\documentclass[onecolumn,10.9pt]{article}
\usepackage[top=1.0in, bottom=1.2in, left=1.2in, right=1.2in]{geometry}
\usepackage{amsmath,amsfonts,amscd,amssymb}
\usepackage{graphicx}
\usepackage{epstopdf}
\usepackage{overpic}
\usepackage{cancel}
\usepackage{rotating}
\usepackage{url}
\usepackage{caption}
\usepackage{color}
\usepackage{rotating}
\usepackage{multirow}
\usepackage{wrapfig}
\usepackage{mathtools}
\usepackage{subeqnarray}
\usepackage{setspace}
\usepackage{palatino} % needs 10
\usepackage[numbers,sort&compress]{natbib}

\usepackage{mathabx}

\usepackage{newtxmath}
\usepackage{subcaption}

\usepackage[bottom,flushmargin,hang,multiple]{footmisc}
\usepackage{lipsum}
\newcommand\blfootnote[1]{%
  \begingroup
  \renewcommand\thefootnote{}\footnote{#1}%
  \addtocounter{footnote}{-1}%
  \endgroup
}

\definecolor{header1}{cmyk}{0,0,0,1}

\title{\LARGE{\vspace{-.55in}\textbf{Data-driven nonlinear aeroelastic models of morphing wings for control}}\vspace{-.175in}}

\author{\normalsize{N. Fonzi$^{1}$, S. L. Brunton$^{2}$ and U. Fasel$^{1*}$}\\
\footnotesize{$^1$ CMASLab, ETH Zurich, 8092 Zurich, Switzerland}\\
\footnotesize{$^2$ Department of Mechanical Engineering, University of Washington, Seattle, WA 98195, United States\vspace{-.2in}}
}
\date{}
\begin{document}
\maketitle

\blfootnote{$^*$ Corresponding author (faselu@ethz.ch).}%\\ \noindent %%%%%%%%%%%%
%%% ABSTRACT
%%%%%%%%%%%%
\begin{abstract}
Accurate and efficient aeroelastic models are critically important for enabling the optimization and control of highly flexible aerospace structures, which are expected to become pervasive in future transportation and energy systems.  
Advanced materials and morphing wing technologies are resulting in next-generation aeroelastic systems that are characterized by highly-coupled and nonlinear interactions between the aerodynamic and structural dynamics.  
In this work, we leverage emerging data-driven modeling techniques to develop highly accurate and tractable reduced-order aeroelastic models that are valid over a wide range of operating conditions and are suitable for control.
In particular, we develop two extensions to the recent dynamic mode decomposition with control (DMDc) algorithm to make it suitable for flexible aeroelastic systems: 1) we introduce a formulation to handle algebraic equations, and 2) we develop an interpolation scheme to smoothly connect several linear DMDc models developed in different operating regimes.  
Thus, the innovation lies in accurately modeling the nonlinearities of the coupled aerostructural dynamics over multiple operating regimes, not restricting the validity of the model to a narrow region around a linearization point. 
We demonstrate this approach on a high-fidelity, three-dimensional numerical model of an airborne wind energy (AWE) system, although the methods are generally applicable to any highly coupled aeroelastic system or  dynamical system operating over multiple operating regimes.
Our proposed modeling framework results in real-time prediction of nonlinear unsteady aeroelastic responses of flexible aerospace structures, and we demonstrate the enhanced model performance for model predictive control.  
Thus, the proposed architecture may help enable the widespread adoption of next-generation morphing wing technologies. 

\vspace{0.2in}

morphing wings

reduced-order modeling

data-driven modeling

dynamic mode decomposition

model predictive control

aeroelastic models

airborne wind energy
\vspace{0.2in}
\end{abstract}

\maketitle

\section{Introduction}
It is expected that highly flexible aeroelastic structures will become ubiquitous in future transportation and energy systems, enabled by advanced materials and emerging morphing wing technologies~\cite{Li2018}.  
Indeed, more responsive and deformable aerodynamic surfaces may have transformative impact in efficiency and maneuverability, as demonstrated by the incredible performance of biological flight systems~\cite{Combes:2001,Birch2001nature,Hedenstrom2007science,song2008aeromechanics}.  
These flexible aerospace structures pose a considerable modeling challenge, as they involve highly coupled and nonlinear interactions between the aerodynamic and structural dynamics.  
Existing aeroelastic models are frequently either linearized about a single operating condition or involve expensive high-fidelity numerical simulations to resolve the relevant spatial and temporal scales~\cite{leishman:06,dowell:2001}.  
In this work, we develop a flexible data-driven modeling architecture, based on the recent dynamic mode decomposition with control (DMDc)~\cite{Proctor2016siads}, to model highly coupled and nonlinear aeroelastic dynamics over the entire flight envelope.  
This work is motivated by a particularly compelling application in airborne wind energy, although the methods discussed are generally applicable to highly coupled aeroelastic systems.  
To achieve these flexible and tractable models, we introduce several innovations to the DMD architecture that may be used more broadly.  
Finally, we demonstrate the efficacy of the resulting models for model predictive control (MPC).

The development and deployment of innovative renewable energy technologies is essential to reduce greenhouse gas emissions. Airborne wind energy (AWE) is a promising technology that extracts power from high-altitude winds using tethered drones. Currently, these drones are equipped with conventional rigid-wings, relying on hinged control surfaces~\cite{Ahrens2014,Cherubini2015}. Replacing the rigid-wings with shape-adaptable or so-called morphing wings has the potential to improve power production by allowing the drone to smoothly adapt to changing flight conditions, thus enabling optimal performance over the full operational regime~\cite{fasel2017}. Such morphing wings are inherently flexible, leading to a tight coupling of their aerodynamic, structural, and rigid-body responses~\cite{Hodges2002,Molinari2014}. This coupling, combined with the vast flight regime such AWE drones operate in, makes modeling and controlling these systems a considerable challenge.

Ground-based power-generator AWE systems, in particular, aim to extract power by periodically reeling-out (traction phase) and reeling-in (retraction phase) the tether. Thus, the drone constantly changes between two distinct operating modes. In the traction phase, maximum power is produced by operating the drone at high flight speeds, large incidence angle, and high-lift forces; whereas in the retraction phase, the load on the tether is minimized by decreasing the incidence angle to reduce the required reel-in power. The drone is therefore required to operate both at high and low wing loading over a wide range of wind speeds, while simultaneously following a desired trajectory, thus creating the need for highly adaptable drones. To design, analyze, and control such a high-dimensional, nonlinear dynamical system, efficient numerical models are needed. Opposed to earlier work on flutter~\cite{wagner:25,Theodorsen:35,karman:38num,sears:41foil,peters:2007} or recent work on high aspect ratio wings~\cite{Patil1998,palacios2010,Afonso2017}, these models need to be valid over a larger set of operational conditions and flight speeds.

A compelling new family of methods capable to approach this problem are emerging data-driven modeling techniques enabling the characterization of such high-dimensional, nonlinear dynamical systems~\cite{Kutz2016book,brunton2019,Brunton2020arfm}. Dynamic mode decomposition (DMD) is a particularly promising technique, enabling the discovery of dynamical systems from high-dimensional data by decomposing complex dynamics into simple representations based on spatiotemporal coherent structures~\cite{Schmid2010jfm,Rowley2009jfm,Tu2014jcd,Kutz2016book}. By using measurements of the system, DMD extracts the dynamics without the need to know the underlying equations, opposed to physics-based models build on established mathematical and physical laws. DMD was first introduced by Schmid~\cite{Schmid2010jfm} and has since gained traction for modeling systems exhibiting nonlinearities. DMD is strongly connected to the Koopman operator, which is an infinite-dimensional linear operator representing nonlinear dynamical systems~\cite{Rowley2009jfm}. Therefore, DMD is a promising candidate to model the nonlinear dynamics inherent to morphing AWE drones.

In this work, we present an efficient and accurate unsteady aeroelastic reduced-order model (ROM) for flexible structures, applicable to AWE morphing wing drones. The method is outlined in figure 1. Two innovations are introduced. First, an extension of the DMD method is developed for algebraic differential equations to generate a reduced-order unsteady aerodynamic and aeroelastic model. To generate the ROM, the algorithm sequentially applies a mode superposition method on a detailed 3-D structural finite element model~\cite{Besselink2013}, followed by using the extended algebraic DMD method on a coupled structural and unsteady 3-D panel method~\cite{Fasel2019rom,katz2001}. Second, an interpolation scheme smoothly connecting multiple ROMs valid at different flight velocities is introduced. This allows accurate faster-than-real-time prediction of the nonlinearities of the coupled aerostructural-dynamics of flexible AWE drones over the full nonlinear flight regime.

\begin{figure}[t]
\vspace{-.25in}
\centering\includegraphics[width=.99\textwidth]{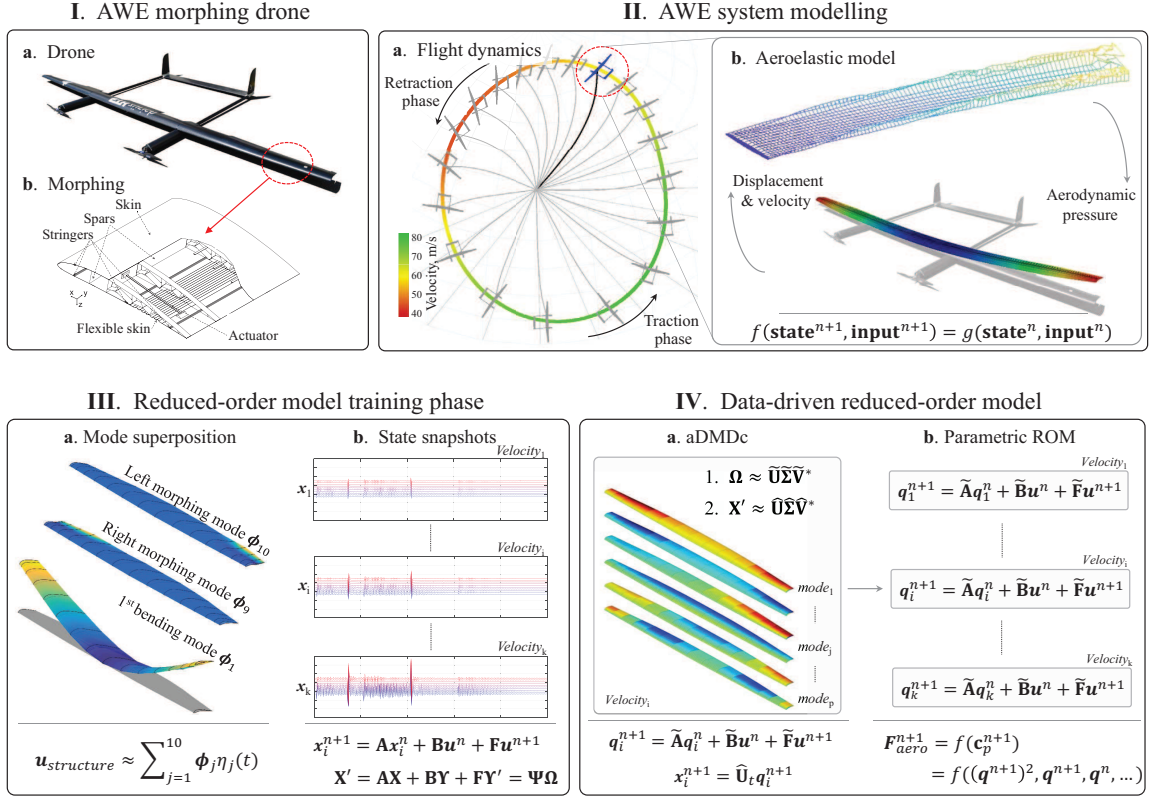}
\vspace{-.1in}
\caption{\textbf{Method overview.} Modeling of AWE morphing drones: \textbf{I}. Exemplary AWE morphing drone (designed and manufactured within the ftero AWE project at ETH Zurich~\cite{ftero,keidel2017ftero}), exploiting camber-morphing for roll-control~\cite{Molinari2014,Keidel2017smasis,Keidel2020}. \textbf{II}. AWE system modeling, consisting of the coupled flight dynamic and aeroelastic models. \textbf{III}. Reduced-order model training phase, consisting of \textbf{a}. extracting the most important structural and morphing modes, \textbf{b}. training the model from snapshots of the states $\boldsymbol{x}$ of the aeroelastic model excited through impulses in the inputs $\boldsymbol{u}$ for a set of flight speeds. \textbf{IV}. Data-driven reduced-order modeling method, consisting of \textbf{a}. generating the parametric algebraic DMD with control (aDMDc) model as in \S~\ref{Sec:Methods} (depicted is the doublet distribution on the wing for a number of aDMDc modes) and \textbf{b}. applying the parametric ROM to calculate the aerodynamic forces and moments $\boldsymbol{F}_{aero}$ generated by the morphing wing for a specific input $\boldsymbol{u}$.}
\label{fig_1}
\end{figure}
\vspace{.1in}

\newpage

\noindent The paper is organized as follows. In Sec. 2, the state-of-the-art in modeling flexible aerospace structures, reduced-order modeling, DMD, and MPC is reviewed. In Sec. 3, the proposed modeling approach is introduced, specifically the extension of DMD for algebraic differential equations and to interpolate between local linear models. In Sec. 4, a number of numerical examples, including a NACA0012 rigid-wing, a morphing AWE wing, and an MPC test case are discussed, highlighting the applicability of the ROM to aerodynamics-only, aeroelastic, and control problems. In Sec. 5, the results and the potential applicability for modeling and controls of general aerospace structures and other dynamic systems is discussed.

\newpage

\section{Background}
\subsection{Modeling of flexible aerospace structures}

The interaction between structural deformations and aerodynamic forces has long been recognised in the field of aeronautic as being of paramount importance. The first wind tunnels at NACA/NASA were specifically dedicated to aeroelastic studies. Early flight suffered from aeroelastic issues, and as flight speed increased, it was not possible to neglect these effects~\cite{fung2008}. Now, it is common practise to consider aeroelastic effects early in the design, to avoid expensive redesigns. Due to the different equations for the structure and for the aerodynamics, the two problems are typically modeled with separate techniques, to be coupled later with an appropriate scheme~\cite{katz2001}. Splines are usually used~\cite{rodden1994} to interpolate the structural displacements onto the aerodynamic grid, and the aerodynamic forces onto the structural nodes.

Nonlinear beam models are commonly used to describe the characteristic of flexible structures with a dominant spatial dimension~\cite{Hodges2002,Patil1998,palacios2010}. The sectional properties of such structures are usually pre-calculated along their dominant direction. However, more refined models are required in the case of morphing and geometrically complex wings that exhibit flexibility in the chordwise direction and, therefore, strongly interact with the aerodynamics. Detailed finite element models, based on both beam and shell elements, are therefore used to accurately represent the characteristics of such structures~\cite{Fasel2019rom}. To increase computational efficiency, model reduction based on modal decomposition techniques are often applied~\cite{Besselink2013,Fasel2019rom}.

In terms of the aerodynamics, different models have been applied to fluid-structure interaction (FSI) problems, depending on the tradeoff between computational cost and accuracy of the simulation. Early studies considered simple 2D geometries with analytical models for the aerodynamics, based on unsteady potential flow theory~\cite{Theodorsen:35}. These methods rapidly evolved, and extensions to 3D problems, based on strip theory, are still used today~\cite{kier2005}. However, it is now common to use the doublet lattice method (DLM)~\cite{albano1969} for the unsteady aerodynamic generalised forces, and its steady counterpart the vortex lattice method (VLM) \cite{falkner1943, hedman1965}, for aeroelastic analysis. Compared to computational fluid dynamics based on the Navier-Stokes equations, these methods are significantly more efficient, and they provide results that are accurate enough for the early design stages. These methods do not require discretising the volume surrounding a body, but instead reduce the problem to an equivalent formulation on the boundaries of the domain, so that only the wing surface must be discretised. Therefore, these methods do not suffer from issues related to mesh deformations~\cite{maute2001}. On the other hand, they do not represent viscous effects and are not suited for transonic applications, with the exception of Morino's method~\cite{morino1974}.

Panel methods are mainly divided into frequency-domain and time-domain methods.  
Representing unsteady aerodynamics in the frequency domain is useful for flutter predictions. A linear state space model is usually preferred for response analysis and control design~\cite{Brunton2013jfs,Brunton2013jfm,Brunton2014jfs}. 
In the field of AWE, the benchmark problem for the method in this paper, researchers have applied several time-domain models, ranging from lifting line methods~\cite{drela1999} to quasi-steady approximations of a 3D panel method, based on source and doublet distributions~\cite{fasel2017}. However, none of these approaches considered the full unsteadiness of the flow. 

Murua et al.~\cite{murua2012} proposed a promising approach based on the unsteady vortex lattice method (UVLM). The UVLM is the direct extension of VLM in the unsteady time domain. In this specific case, linearising the nonlinear equations for lifting surfaces and aerodynamics, the state-space form is easily obtained. Additionally, it is straightforward to include the flexible-body dynamics in the simulation~\cite{palacios2010}.
In practise, the full equations of the UVLM can be summarised as:
\begin{subequations}
\begin{align}
    \label{eq:UVLMa}
    f(\boldsymbol{x}^{n+1},\boldsymbol{u}^{n+1}) &= g(\boldsymbol{x}^{n},\boldsymbol{u}^{n})\\
    \label{eq:UVLM}
    \boldsymbol{y}^n &= h(\boldsymbol{x}^{n},\boldsymbol{u}^{n}),
\end{align}
\end{subequations}
where $f$, $g$ and $h$ are general nonlinear functions, $\boldsymbol{x}$ is the vector of states, composed of structural and aerodynamic nodes positions, $\boldsymbol{u}$ is the vector of inputs (e.g. the flight condition) and $\boldsymbol{y}$ is the vector of outputs (the aerodynamic forces). The superscript indicates the time step.
Note that these equations depend on the control inputs at the next time step as a result of the impulsive part of the aerodynamics. 
Usually, when artificial aerodynamic states are added and a state-space form is obtained, this is reflected in the feed-trough matrix to the outputs. However, if the correct evolution of the states must be considered, the immediate effect of the inputs must be modeled.

Under the assumption of small displacements around the trim condition, thus the deformed equilibrium, the equations can be linearized, resulting in a state space form that can be coupled with the structural equations:

\begin{subequations}
\begin{align}
    \textbf{E}\boldsymbol{x}^{n+1} + \textbf{F}\boldsymbol{u}^{n+1} = \textbf{A}\boldsymbol{x}^{n} + \textbf{B}\boldsymbol{u}^{n}\\
       \boldsymbol{y}^n = \textbf{C}\boldsymbol{x}^{n} + \textbf{D}\boldsymbol{u}^{n}.
\end{align}
\end{subequations}

The limitations of UVLM are related to the approximation of the surface as an infinitely thin sheet. If the camber effect is important, as in the context of morphing structures or flexible airfoils, this requires more detailed models, such as the unsteady 3D panel method based on both sources and doublets~\cite{katz2001}. In this case, the equations for the aerodynamics are:

\begin{equation}
    \textbf{A}_b\boldsymbol{\mu}_b^{n+1} + \textbf{A}_w\boldsymbol{\mu}_w^{n+1} + \textbf{B}\boldsymbol{w}^{n+1} = 0,
\end{equation}
where $\textbf{A}_b$ is the aerodynamic influence matrix of the body, $\textbf{A}_w$ is the wake influence matrix, $\boldsymbol{\mu}$ is a vector containing the doublet strengths, and $\boldsymbol{w}$ is the downwash at the control points.
The wake node positions are then evolved with a simple advection equation, using the local velocity.
These equations are similar to \ref{eq:UVLMa}, and the computation of the forces are governed by a similar expression.
Both the UVLM and the 3D panel methods involve algebraic-differential equations.  It is essential that our ROMs respect this structure. 

\vspace{0.2in}

\subsection{Reduced-order aeroelastic models}

The methods above provide accurate models of aeroelastic effects.  However, faster models are often necessary for optimisation and control, even at the expense of some fidelity.  
This tradeoff between accuracy and efficiency has motivated reduced-order models, which model the behaviour of the system with as few states as possible.
There is a wide variety of model reduction techniques in the literature~\cite{brunton2019}. Both data-driven and semi-analytical approaches are common. The latter is exemplified by the modal reduction of structural dynamics: an eigenvalue problem is solved and a reduced set of orthogonal modes are used to describe the state of the system. Other analytical methods project the governing equations onto a set of data-driven modes~\cite{lieu2005,amsallem2008}, for example obtained via the proper orthogonal decomposition (POD); it maybe possible to develop parametric models, although this procedure may be cumbersome.  

In recent years, data-driven approaches have become increasingly powerful and widely adopted.  
Many data-driven modeling techniques may be categorized as system identification, where a model for the input--output behavior is constructed based on data.  
Time domain techniques are especially common, such as the aerodynamic impulse response (AIR)~\cite{silva2001,guendel2001,silva2001.2,raveh2001}, where the system is perturbed with an impulsive input, and the output response may be used to predict the response to future input maneuvers via convolution.  Although these techniques are typically linearized, nonlinear kernels may also be employed in a Volterra series~\cite{Brockett1976automatica,guendel2001}. 
State-space realizations are becoming increasingly common, especially for control applications~\cite{Brunton2013jfs,Brunton2014jfs}.  
The eigensystem realization algorithm~\cite{juang1985} was developed specifically to model the structural response of aerospace structures, although the resulting model is formulated in terms of a nonphysical state that is difficult to interpret.  
However, when the entire physical state can be observed, it is possible to construct state-space models in terms of a reduced state that is related to the physical domain.  
The dynamic mode decomposition (DMD)~\cite{Schmid2010jfm,Rowley2009jfm,Tu2014jcd,Kutz2016book} and the sparse identification of nonlinear dynamics (SINDy)~\cite{Brunton2016pnas,Loiseau2017jfm,Loiseau2018jfm} both result in physically interpretable models of the dynamics, and they have been extended to input--output systems for control~\cite{Proctor2016siads,Kaiser2018prsa}. 

DMD results in a linear time-invariant (LTI) system, which is valid in a particular operating regime.  
There are several approaches to interpolate LTI systems to obtain a linear parameter varying (LPV) system, as a function of the flight condition~\cite{de2009, benner2015, baur2017, balas2002,Hemati2016aiaa}. However, it is generally challenging to interpolate models; indeed, the direct interpolation of the matrices only works for small systems~\cite{steinbuch2003}, but for large systems the eigenvalues of the system become unstable~\cite{wassink2005}. If the transfer function is interpolated, assuming oscillatory movement, this problem is solved, but there are still limitations related to the change of an oscillatory couple of poles, to two real-valued poles~\cite{de2009,paijmans2008}. In general, there is no universal solution for the interpolation of LTIs, and this work will develop an interpolation scheme specifically for DMD models.

\vspace{0.4in}

\subsection{Dynamic mode decomposition with control}
In this work, we will develop an interpolation scheme to combine local models that are obtained via dynamic mode decomposition with control (DMDc)~\cite{Proctor2016siads}. There are several advantages to DMDc over other linear state-space modeling techniques, primarily the formulation of the models in terms of physically interpretable modes~\cite{Kutz2016book}.  Further, it has been shown by Kaiser et al.~\cite{Kaiser2018prsa} that DMDc models may be nearly as effective as the full nonlinear dynamics for model predictive control, even in chaotic systems.  

In DMDc, it is assumed that the entire state and all control inputs may be observed to train the model, and the evolution of this state may be expressed as a linear system 
\begin{equation}
\boldsymbol{x}^{n+1} = \textbf{A} \boldsymbol{x}^{n} + \textbf{B} \boldsymbol{u}^n,
\label{eq:dmd}
\end{equation}
where $\textbf{A}$ and $\textbf{B}$ are unknown constant matrices, $\boldsymbol{x}$ is the state vector, and $\boldsymbol{u}$ is the input vector.

The unknown matrices $\textbf{A}$ and $\textbf{B}$ are solved through a regression procedure, based on time-series measurement data:
\begin{subequations}
\begin{align*}
\textbf{X} &= [\ \boldsymbol{x}^{1}\  |\  \boldsymbol{x}^{2}\  | \ \boldsymbol{x}^{3}\  | \ ...\  | \ \boldsymbol{x}^{m-1}\ ]\\
\textbf{X}' &= [\ \boldsymbol{x}^{2}\  |\  \boldsymbol{x}^{3}\  | \ \boldsymbol{x}^{4}\  | \ ...\  |\  \boldsymbol{x}^{m}\ ]\\
\boldsymbol{\Upsilonup} &= [\ \boldsymbol{u}^{1}\  |\  \boldsymbol{u}^{2}\  | \ \boldsymbol{u}^{3}\  | \ ...\  | \ \boldsymbol{u}^{m-1}\ ].
\end{align*}
\end{subequations}
The original dynamics may be written in terms of these data matrices as
\begin{equation}
\textbf{X}' \approx \textbf{A} \textbf{X} + \textbf{B} \boldsymbol{\Upsilonup} = \boldsymbol{\Psiup} \boldsymbol{\Omegaup}
\label{eq:dmd3}
\end{equation}
where $\boldsymbol{\Psiup} = [\textbf{A}\ \textbf{B}]$ and $\boldsymbol{\Omegaup} = [\textbf{X}^T\ \boldsymbol{\Upsilonup}^T]^T$. 
The matrix of known terms $\boldsymbol{\Omegaup}$ can be approximated by a singular value decomposition (SVD) as $\boldsymbol{\Omegaup} = \textbf{U} \boldsymbol{\Sigmaup} \textbf{V}^{T}$.  This expression may be truncated, taking into account only the vectors associated with the largest elements of $\boldsymbol{\Sigmaup}$. This gives the approximation $\boldsymbol{\Omegaup} \approx \textbf{U}_{t} \boldsymbol{\Sigmaup}_{t} \textbf{V}_{t}^{T}$, where the subscript $t$ denotes truncated matrices. It is then possible to solve for the unknown terms in $\boldsymbol{\Psiup}$ via the pseudo-inverse of $\boldsymbol{\Omegaup}$:
\begin{equation}
\boldsymbol{\Psiup} \approx \textbf{X}' \textbf{V}_{t} \boldsymbol{\Sigmaup}_{t}^{-1} \textbf{U}_{t}^{T}.
\end{equation}
It is possible to build a reduced-order model by projecting the dynamics onto the leading modes $\hat{\textbf{U}}_t$ from the SVD of $\textbf{X}'\approx \hat{\textbf{U}}_t\hat{\boldsymbol{\Sigmaup}}_t\hat{\textbf{V}}_t^T$:
\begin{equation}
\tilde{\boldsymbol{\Psiup}} = \hat{\textbf{U}}_{t}^{T} \boldsymbol{\Psiup} \hat{\textbf{U}}_{t}.
\label{eq:redA}
\end{equation}
The resulting $\tilde{\boldsymbol{\Psiup}}$ matrix can then be split into $\tilde{\textbf{A}}$ and $\tilde{\textbf{B}}$, so that:
\begin{equation}
    \boldsymbol{q}^{n+1} = \tilde{\textbf{A}} \boldsymbol{q}^{n} + \tilde{\textbf{B}} \boldsymbol{u}^{n},
\end{equation}
where $\boldsymbol{x} = \hat{\textbf{U}}_{t} \boldsymbol{q}$, and $\boldsymbol{q}$ is the vector of mode amplitudes.

\vspace{0.4in}

\subsection{Model predictive control}

In this section, we introduce the MPC problem, outline its potential benefits, and discuss the importance of developing efficient ROMs to allow real-time control. MPC is an effective model-based control, which has revolutionized the industrial control landscape~\cite{camacho2013model}, as it enables the control of strongly nonlinear systems with constraints, which are difficult to handle using traditional linear control approaches~\cite{garcia1989model,morari1999model,mayne2014automatica}. 

MPC computes a control sequence  $\boldsymbol{u}(\boldsymbol{x}_j) = \{\boldsymbol{u}_{j+1},...,\boldsymbol{u}_{j+m_c}\}$, given a measurement $\boldsymbol{x}_j$, by solving a constrained optimization over a receding horizon $T_c = m_c \Delta t$. 
At each time step, the optimization is repeated, updating the control sequence over the control horizon and applying the first control action $\boldsymbol{u}_{j+1}$ to the system. The optimal control sequence $\boldsymbol{u}(\boldsymbol{x}_j)$ is obtained by minimizing a cost function $J$ over a prediction horizon $T_p = m_p \Delta t \geq T_c$. The cost function is 
\begin{equation}
    J(\boldsymbol{x}_j) = \sum_{k=0}^{m_p-1} \parallel \hat{\boldsymbol{x}}_{j+k} - \boldsymbol{x}_{k}^* \parallel_{\textbf{Q}}^2 + \sum_{k=1}^{m_c-1} (\parallel \hat{\boldsymbol{u}}_{j+k} \parallel_{\textbf{R}_u}^2 + \parallel \Delta \hat{\boldsymbol{u}}_{j+k} \parallel_{\textbf{R}_{\Delta u}}^2 )\,, 
\end{equation}
subject to the discrete-time dynamics and other constraints. The cost function thus penalizes deviations of state from the reference trajectory, the control expenditure, and the rate of change of the control signal, with each term weighted by the matrices  $\textbf{Q}$, $\textbf{R}_u$, and $\textbf{R}_{\Delta u}$, respectively. 
To enable this optimization loop to run in real-time on a flight controller, MPC relies on efficient models and high-performance computing. In AWE and general flight systems, the control must rapidly respond to disturbances on short time-scales, as time delays from sensors, signal transduction, or processing can destroy the robustness of feedback control, putting limitations on the achievable performance~\cite{Doyle2013book}. Currently, MPC is not well suited for control of such complex, high-dimensional system with fast timescales, especially with power-constrained or computationally limited hardware available on state-of-the-art AWE drones or lightweight flexible aircraft. The short timescales associated with agile flight in a complex unsteady fluid environment make it challenging to arrive at control decisions with the small latency required for robust performance. 
Gust disturbances are very rapid, and computing and estimating the dynamics of the unsteady fluid requires considerable computational resources. Therefore, instead of a detailed model-based feedback control strategy that spans all relevant timescales, it is essential to develop ROMs that balance accuracy and efficiency.

The benefit of MPC lies in simple and intuitive tuning and the ability to control complex phenomena, including systems with time delays, non-minimum phase dynamics, and instability. It is also  simple to include known constraints and multiple operating conditions, and it provides the flexibility to formulate and tailor a control objective. 
The major challenge of MPC lies in the development of a suitable model via existing system identification or model reduction techniques~\cite{Brunton2015amr,Kaiser2018prsa}, which may require expensive and time-consuming data collection and computations. 
Nonlinear models based on machine learning, such as neural networks, are increasingly used due to advances in computing power, and recently deep reinforcement learning has been combined with MPC~\cite{Peng.2009,Zhang2016icra}, yielding impressive results in the large-data and high-performance computing limit. 
However, many modern techniques in machine learning (e.g., neural networks) rely on access to massive data sets, have limited ability to generalize, do not readily incorporate known physical constraints, and require expensive and time-consuming computations. 
Instead, Kaiser et al.~\cite{Kaiser2018prsa} showed that a simple linear model obtained via dynamic mode decomposition with control (DMDc)~\cite{Proctor2016siads} performs nearly as well with MPC as a full nonlinear model, and may be trained in a surprisingly short amount of time. 

Currently, several studies have applied MPC to AWE, although they rely on simplified models~\cite{Ilzhofer2007,Gros2013,Lucia2014,Wood2017,Diwale2017,Stastny2019}. For flexible aerospace structures~\cite{Haghighat2012,Simpson2014,Wang2016,Wang2018,Artola2019}, most control strategies rely on linearised models valid only around a single steady state trim position or on successive linearisation of the underlying system.  
These models do not allow for  optimal control actions over a large range of flight speeds. Thus, there is a need to develop fast and accurate ROMs covering a large range of flight speeds, which can  be used for MPC.

\section{Methods}\label{Sec:Methods}
The models and methods used for the current research are presented in this section. The full structural model is based on a linear structure, composed of beams and shells, while the aerodynamics are modeled using an unsteady 3D panel method. The coupling is obtained via a thin plate spline (TPS)~\cite{harder1972} and inverse distance weighting (IDW) interpolation~\cite{shepard1968} and the integration in time is performed using the Newmark method~\cite{newmark1959} with a tight coupling between structural part and aerodynamic model. The full model is described in detail in the Appendix.

Taking inspiration from the approach of~\cite{murua2012}, we develop a time-domain ROM.  However, instead of a mathematically formal linearisation, we leverage the data-driven DMDc~\cite{Proctor2016siads}. This method has been applied to a wide variety of problems, ranging from fluid dynamics to disease modeling~\cite{Kutz2016book,brunton2019}, and it has shown promise in the model predictive control of complex systems~\cite{Kaiser2018prsa}. 
In the following, we develop extensions to DMDc to handle algebraic-differential systems and to interpolate between multiple operating regimes, extending the applicability of these methods to aeroelastic systems over a large operating regime.

\subsection{Algebraic dynamic mode decomposition method aDMDc}
Here we extend the DMDc algorithm to handle algebraic-differential equations, as are found in aeroelastic problems:  
\begin{equation}
f(\textbf{x}^{n+1},\textbf{u}^{n+1}) = g(\textbf{x}^{n},\textbf{u}^{n}).
\label{eq:systemgoverning}
\end{equation}
We seek a linear system of the form:
\begin{equation}
\textbf{q}^{n+1} = \tilde{\textbf{A}} \textbf{q}^{n} + \tilde{\textbf{B}} \textbf{u}^{n} + \tilde{\textbf{F}} \textbf{u}^{n+1}.
\label{eq:dmd+control+}
\end{equation}
Therefore, we will modify the standard DMDc to include the $\mathbf{F}$ matrix. The modification of the system is straightforward. In addition to the standard DMDc data matrices, we include the matrix of inputs shifted in time:
\begin{equation}
\label{eq:dmd+control+3}
\boldsymbol{\Upsilonup}' = [\ \textbf{u}^{2}\  |\  \textbf{u}^{3}\  | \ \textbf{u}^{4}\  | \ ...\  |\  \textbf{u}^{m}\ ].
\end{equation}
Again, we define:

\begin{equation}
\label{eq:dmd+control+4}
\boldsymbol{\Psiup} = \begin{bmatrix} \textbf{A}\\ \textbf{B}\\ \textbf{F}\end{bmatrix}\ \ \ \ \ \ \ \boldsymbol{\Omegaup} = \begin{bmatrix} \textbf{X} \\ \boldsymbol{\Upsilonup} \\ \boldsymbol{\Upsilonup}'\end{bmatrix}.
\end{equation}
Thus, it is possible to write the dynamics as:
\begin{equation}
\label{eq:dmd+control+5}
\textbf{X}' \approx \boldsymbol{\Psiup}\boldsymbol{\Omegaup}.
\end{equation}
In order to solve for $\boldsymbol{\Psiup}$, we compute the SVD of $\boldsymbol{\Omegaup}$:
\begin{equation}
\label{eq:dmd+control+6}
\boldsymbol{\Omegaup} = \textbf{U}\boldsymbol{\Sigmaup} \textbf{V}^{T},
\end{equation}
and invert $\boldsymbol{\Omegaup}$ via the pseudo-inverse, as with DMDc. 
As before, we truncate the SVD using the optimal hard threshold criteria of Gavish and Donoho~\cite{gavish2014}. 
We then obtain:

\begin{subequations}
\label{eq:dmd+control+9}
\begin{align*}
\tilde{\textbf{A}} &= \hat{\textbf{U}}_{t}^T\textbf{X}'\textbf{V}_{t}\boldsymbol{\Sigmaup}^{-1}_{t}\textbf{U}^{T}_{1}\hat{\textbf{U}}_{t}\\
\tilde{\textbf{B}} &= \hat{\textbf{U}}_{t}^T\textbf{X}'\textbf{V}_{t}\boldsymbol{\Sigmaup}^{-1}_{t}\textbf{U}^{T}_{2}\\
\tilde{\textbf{F}} &= \hat{\textbf{U}}_{t}^T\textbf{X}'\textbf{V}_{t}\boldsymbol{\Sigmaup}^{-1}_{t}\textbf{U}^{T}_{3}.
\end{align*}
\end{subequations}
The matrices $\textbf{U}_1$, $\textbf{U}_2$, and $\textbf{U}_3$ are obtained by splitting  the truncated SVD $\textbf{U}_t$ according to the structure of $\boldsymbol{\Omegaup}$ in Eq.~\eqref{eq:dmd+control+4}, as in DMDc, and $\hat{\textbf{U}}_t$ are the leading SVD modes of $\textbf{X}'\approx \hat{\textbf{U}}_t\hat{\boldsymbol{\Sigmaup}}_t\hat{\textbf{V}}_t^T$. 
This procedure, results in a data-driven approximation of the system in  Eq.~\eqref{eq:dmd+control+}.  It is possible to go between the reduced state $\textbf{q}$ and the full physical state $\textbf{x}$ via the modes $\hat{\textbf{U}}_{t}$.

The procedure to generate the ROM is summarized in Algorithm 1.

\begin{table}[t]
\label{table_example}
\begin{tabular}{llr}
\hline
\multicolumn{3}{l}{{\bf Algorithm 1:} Generate aDMDc - ROM for differential-algebraic system of equations} \\
\hline
\multicolumn{3}{l}{{\bf Input:} Snapshot matrices $\textbf{X, X'}, \boldsymbol{\Upsilonup}, \boldsymbol{\Upsilonup}'$} \\
1: & {\bf procedure} aDMDc($\textbf{X, X'}, \boldsymbol{\Upsilonup}, \boldsymbol{\Upsilonup}'$) & \\
2: & \quad \(\boldsymbol{\Omegaup}= [\textbf{X} \ \boldsymbol{\Upsilonup} \ \boldsymbol{\Upsilonup}']\)  & $\smalltriangleright$ Construct the input space matrix $\boldsymbol{\Omegaup}$ \\
3: & \quad  \(\boldsymbol{\Omegaup} \approx \tilde{\textbf{U}}  \tilde{\boldsymbol{\Sigmaup}}  \tilde{\textbf{V}}^* \) & $\smalltriangleright$ Compute and truncate the SVD of $\boldsymbol{\Omegaup}$ \\
4: & \quad  \(\textbf{X}' \approx \hat{\textbf{U}}  \hat{\boldsymbol{\Sigmaup}}  \hat{\textbf{V}}^* \)
 & $\smalltriangleright$ Compute and truncate the SVD of the output space \textbf{X}' \\
5: & \quad  \(\tilde{\textbf{A}} = \hat{\textbf{U}}^*\textbf{X}'\tilde{\textbf{V}}\tilde{\boldsymbol{\Sigmaup}}^{-1}\tilde{\textbf{U}}^{*}_{1}\hat{\textbf{U}} \) & $\smalltriangleright$ Compute the approximation of the operators \textbf{A, B, F} \\
  & \quad \(\tilde{\textbf{B}} = \hat{\textbf{U}}^*\textbf{X}'\tilde{\textbf{V}}\tilde{\boldsymbol{\Sigmaup}}^{-1}\tilde{\textbf{U}}^{*}_{2} \) & \\
  & \quad \(\tilde{\textbf{F}} = \hat{\textbf{U}}^*\textbf{X}'\tilde{\textbf{V}}\tilde{\boldsymbol{\Sigmaup}}^{-1}\tilde{\textbf{U}}^{*}_{3} \) & \\
6: & \quad {\bf return} $\tilde{\textbf{A}}$, $\tilde{\textbf{B}}$, $\tilde{\textbf{F}}$, $\hat{\textbf{U}}_t$ & $\smalltriangleright$ Return the ROM and the modal matrix \\
7: & {\bf end procedure} & \\\hline
\end{tabular}
\vspace*{-4pt}
\end{table}

\subsection{aDMDc for flexible aerospace structures}
In this research, we will apply the aDMDc algorithm from above to aeroelastic systems.  
Instead of developing separate ROMs for the aerodynamic and structural models, we design a single monolithic ROM for the coupled aeroelastic system.   
This approach has the benefit of reducing the dimension of the external inputs to the system.  If we developed two isolated models, then all of the structural modes would be inputs to the aerodynamic system.  Instead, in our coupled model, only the terms influencing the flight condition of the body are inputs to the system:

\begin{equation}
\label{eq:dmd+controlinput}
\textbf{u} = \begin{bmatrix} \alpha & p & q & r & F_{1} & \dots & F_{i} & \dots & F_{k} \end{bmatrix}^T,
\end{equation}
where $\alpha$ is the angle of attack, $p,\ q,$ and $r$ are the roll, pitch, and yaw rotation rates and $F_{i}$ is a general normalised actuation force (with values from 0 to 1) required for the morphing wing actuation inputs. The sideslip angle is not included as an input because the effect was considered negligible, but it can easily be added.

Before applying aDMDc, we conduct a modal analysis to identify the most important structural modes required to describe a general motion. The ROM is then developed in terms of these reduced structural modes and the aerodynamics. The state $\textbf{x}$ of the system is then:

\begin{equation}
\label{eq:dmd+controlstate}
\textbf{x} = \begin{bmatrix} \boldsymbol{\eta} \\ \boldsymbol{\mu}\end{bmatrix},
\end{equation}
where $\boldsymbol{\eta}$ contains the amplitudes of the structural modes that are retained and $\boldsymbol{\mu}$ contains the doublet strengths. Only the doublets required to compute the aerodynamic forces are retained, thus only those on the lifting surface and not those of the wake.
The data-driven method is trained using impulsive inputs, although it is possible to use other input sequences. 

If the total forces on the body are of interest, for example if the aeroelastic system must be included in a multibody framework, or we want to study the stability and control derivatives, the distribution of coefficient of pressure can be computed with the full nonlinear expression, described in more detail in the appendix. Indeed, the relation between doublet distribution and pressure distribution is a nonlinear relation and, in order to retain accuracy, it can be used to obtain the forces. The procedure to generate the ROM is summarized in Algorithm 2.

\begin{table}[t]
\label{table_example}
\begin{tabular}{llr}
\hline
\multicolumn{3}{l}{{\bf Algorithm 2:} Hybrid aDMDc and structural mode superposition method} \\
\hline
\multicolumn{3}{l}{{\bf Input:} Mass and stiffness matrix \textbf{M, K}, flight state $V_{\infty}$} \\
1: & {\bf procedure} $\text{aDMDc}_{\text{AE}}$(\textbf{M, K}, $V_{\infty}$) & \\
2: & \quad $\text{eig}(\textbf{M}\ddot{\boldsymbol{x}} + \textbf{K}\boldsymbol{x} = 0)$  & $\smalltriangleright$ Get ROM from mode superposition of structure \\
3: & \quad $\text{impulse}(\alpha, p, q, r, F_1, \dots, F_k)$  & $\smalltriangleright$ Train model and generate snapshots \textbf{X, X', U, U'} \\
4: & \quad  aDMDc & $\smalltriangleright$ Run Algorithm 1 \\
5: & \quad {\bf return} $\tilde{\textbf{A}}$, $\tilde{\textbf{B}}$, $\tilde{\textbf{F}}$, $\hat{\textbf{U}}_{t}$ & $\smalltriangleright$ Return the ROM and the modal matrix \\
6: & {\bf end procedure} & \\\hline
\end{tabular}
\vspace*{-4pt}
\end{table}

\subsection{Parametric aDMDc}
\begin{figure}[t]
\vspace{-.5in}
	\centering   
	\includegraphics[width=0.9\textwidth]{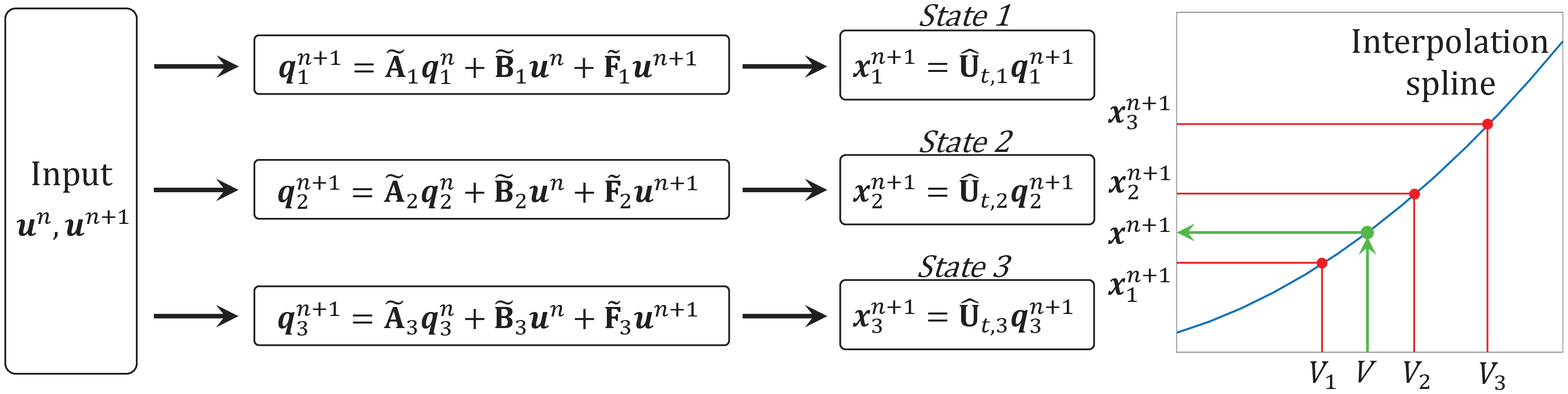}
    \vspace{-.1in}
	\caption{Interpolation of three different ROMs, working at three different velocities (red), to obtain the output $\boldsymbol{x}^{n+1}$ at an intermediate value of velocity $V$ (green).}
	\label{fig:interpolation}
	\vspace{-.1in}
\end{figure}

The method above is able to reproduce the entire aeroelastic system at one flight velocity. Thus, we will construct a separate aDMDc ROM for each velocity condition.
For our AWE application, it is necessary to interpolate multiple ROMs across a wide range of flight conditions.  
There are several valid approaches to interpolate multiple LTI systems into an LPV system.  
Due to the large state dimension, we opt not to interpolate the high-dimensional model, as it is likely to result in unstable models. 
Instead, because the low-dimensional aDMDc model state $\textbf{q}$ may be used to reconstruct the high-dimensional state $\textbf{x}$, we run multiple local ROMs independently and then interpolate the high-dimensional state using a spline method.  
This is a distinct advantage of the DMD approach over other subspace identification methods.  
This approach bypasses the issues with standard interpolation schemes, related to the assumption of slow varying scheduling parameter~\cite{benner2015}. A graphical explanation of this procedure is shown in figure \ref{fig:interpolation}.

\section{Results}

In this section, several applications of the proposed method are outlined. First, we present a simple example illustrating the need for the modified aDMDc algorithm. Next, we demonstrate the approach to model the aerodynamics on a rigid NACA0012 wing.  We then extend this approach to challenging aeroelastic morphing wing, both at a single flight condition and interpolated across several flight conditions.   Finally, the aDMDc model is used for MPC, demonstrating the ability to effectively control a nonlinear system with constraints.

\subsection{Generic differential-algebraic system of equations}

The addition of the $\textbf{F}$ matrix is important to retain the stability of the system. This is better understood with the following example, assuming known dynamics of the form:
\begin{equation}
\begin{bmatrix} q_{1}^{n+1} \\ q_{2}^{n+1} \end{bmatrix} =\begin{bmatrix} 0.1 & 0 \\ 0 & 0.5 \end{bmatrix} \begin{bmatrix} q_{1}^{n} \\ q_{2}^{n} \end{bmatrix} + \begin{bmatrix} 2 \\ 3 \end{bmatrix} u^{n} +\begin{bmatrix} 10 \\ 1.5 \end{bmatrix}u^{n+1}.
\label{eq:dmd+controlex}
\end{equation}
Applying the standard DMDc algorithm, with only the $\textbf{A}$ and $\textbf{B}$ matrices, we obtain:
\begin{equation}
\textbf{A} = \begin{bmatrix} -1.2876 & 1.6203 \\ -0.2081 & 0.7430 \end{bmatrix},\ \ \ \ \ \ \ \ \ \ \ \textbf{B} = \begin{bmatrix} 15.6196 \\ 5.0429 \end{bmatrix}.
\label{eq:dmd+controlex2}
\end{equation}
The resulting dynamics are not only incorrect, but they are unstable.  In contrast, the aDMDc method, results in the correct, stable dynamics of the system: 
\begin{equation}
\textbf{A} = \begin{bmatrix} 0.1 & 0 \\ 0 & 0.5 \end{bmatrix}\ \ \ \ \ \ \ \ \ \ \ \textbf{B} = \begin{bmatrix} 2 \\ 3 \end{bmatrix}\ \ \ \ \ \ \ \ \ \ \ \textbf{F} = \begin{bmatrix} 10 \\ 1.5 \end{bmatrix}.
\label{eq:dmd+controlex3}
\end{equation}
 Note that this system consists of only two states, thus there are no truncation or projections involved in the procedure. However, this first example shows the necessity of adding the $\textbf{F}$ matrix to correctly capture the dynamics of the algebraic-differential system.

\subsection{NACA0012 rigid wing}

We now apply aDMDc to model the aerodynamics of a rigid, unswept, untappered, planar NACA 0012 wing with chord $c=0.4\ $m and aspect ratio $AR = 10$. The wing surface is divided into trapezoidal panels, with 98 panels along the span and 38 along the chord. The discretisation is defined so that no panels are present with high aspect ratio and the results are mesh independent.

To initialize the simulation, the wing is set into motion with a constant speed of $V_\infty = 50$ m/s, a constant angle of attack of $\alpha_0 = 2^\circ$ and no rotation rates. When the steady state is reached, the training phase for the ROM is initiated. In this rigid airfoil case, the system state only consists of the doublet strengths on the surface of the body.
A set of impulsive inputs in the  $\alpha$, $p$, $q$, and $r$ are provided in a random order. The impulse amplitudes are $\Delta \alpha = 8^\circ$ and $\Delta p = \Delta q = \Delta r = 30^\circ/\text{s}$, which are large enough to excite the dynamics, but small enough that the system may be considered as a linearisation around a single flight condition.  
Both positive and negative impulses are used. 

Algorithm 1 is used to obtain a ROM, using the snapshots matrices of the states and inputs recorded in the training phase. For a first comparison, this ROM is used to reproduce the behaviour of the system during the training phase. 
Afterwards, two different testing phases are performed. In these phases, the ROM and the full model are compared. The first test set consists of a sinusoidal roll rate with reduced frequency of 0.1 and amplitude $\Delta p = 18^\circ/\text{s}$ and a sinusoidal change in the angle of attack with reduced frequency of 0.2 and amplitude of $\Delta \alpha = 4^\circ$.
The second test set consists of  a random sequence of impulsive inputs, with the same amplitude as the training phase, but with different order. This case is challenging, as all frequencies are excited by the impulses, fully exploring the limitations of the ROM.

\begin{figure}[t]
\vspace{-.3in}
\begin{center}
    \includegraphics[width=0.9\textwidth]{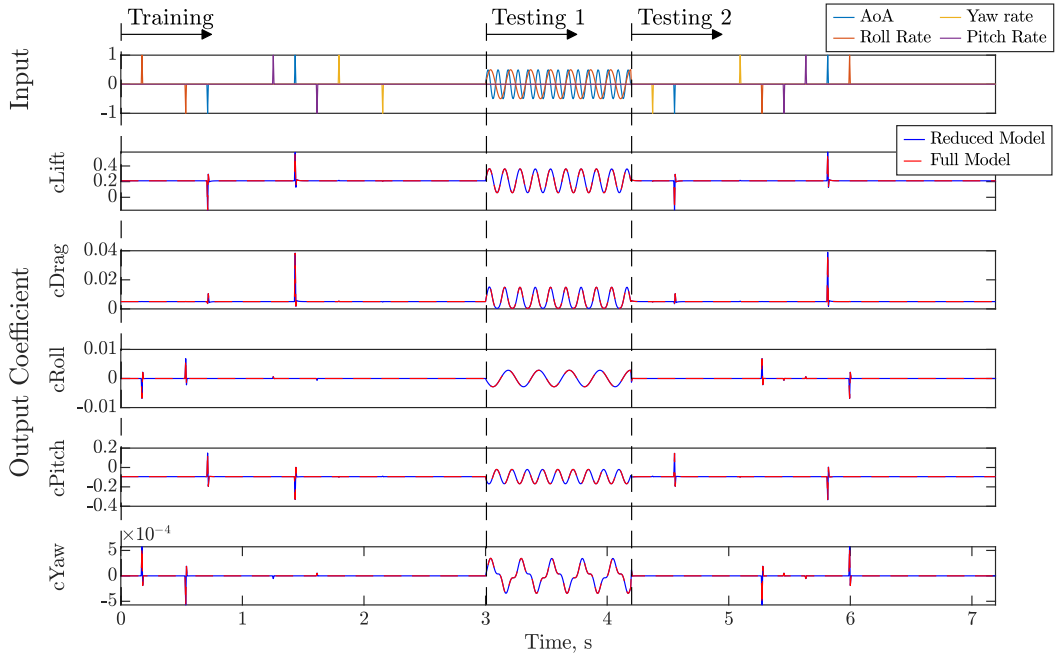}
    \vspace{-.1in}
    \caption{NACA0012 wing aerodynamics-only: training phase followed by testing phases, comparing the full model with the aDMDc model.}
  \label{fig_2}
  \end{center}
  \vspace{-.1in}
\end{figure}

The aerodynamic forces and moment coefficients acting on the wing, for all the three phases are shown in figure~\ref{fig_2}.
The agreement between the full and reduced-order model is excellent.   Importantly, ROM is three orders of magnitude more efficient than the full model, as highlighted in Tab.~\ref{table_example1}, which shows the relative errors and the speed-up factor for the two testing phases.

\begin{table}[b]
\vspace{-.05in}
\caption{NACA0012 ROM speed-up factor $S$ and mean relative errors $R_i$.}\label{table_example1}
\begin{center}
\vspace{-.25in}
\begin{tabular}{lllllll}
\hline
Input: & $S$ & $R_{Lift}$ & $R_{Drag}$ & $R_{Roll}$ & $R_{Pitch}$ & $R_{Yaw}$ \\
\hline
1. Reference: & 2598 & 0.25\% & 0.28\% & 0.25\% & 0.42\% & 0.50\% \\
2. Random: & 2598 & 0.18\% & 0.40\% & 0.23\% & 0.31\% & 0.48\% \\
\hline
\end{tabular}
\vspace*{-4pt}
\end{center}
\vspace{-.35in}
\end{table}

\subsection{Morphing AWE wing}
In this example, we use aDMDc to model a coupled aeroelastic system, given by a flexible and highly cambered wing. The wing is the result of studies performed in the context of the ftero project at ETH Zurich~\cite{ftero, keidel2017ftero}. The wing is composed of a CFRP skin and an internal structure based on a Voronoi tessellation. The trailing edge of both the right and left sides of the wing can morph, increasing or decreasing the camber, thus replacing conventional ailerons.

For flexible structures, it is first necessary to reduce the number of states characterising the wing structure, before training the aDMDc ROM. The full finite element morphing wing model is generated using the commercial software Nastran. The mass and stiffness matrix of the structure is extracted and a modal decomposition is performed.  We retain only the first 8 structural modes. Two additional modes are added corresponding to the deformation modes due to morphing. Afterwards, the aerodynamic mesh is generated, as in the previous example by discretizing the surface with trapezoidal panels.  We then construct the ROM following Algorithm 1. In this case, two additional inputs are included to actuate the morphing surfaces, and the state of the system consists of the doublet strengths on the surface of the body and the structural modal amplitudes. 

The linearisation point is the same as before: $V_\infty = 50$ m/s, $\alpha_0 = 0^\circ$ and $p_0 = q_0 = r_0 = 2^\circ$/s. The same set of impulses for $\alpha$, $p$, $q$ and $r$ is used, with the same amplitudes, resulting in a single ROM that is sufficient at the given velocity. In addition, the morphing wing actuation is impulsively fully actuated during the training phase. Two testing phases are considered, as before, consisting of sinusoidal and impulsive inputs.  

\begin{figure}[t]
\vspace{-.3in}
\begin{center}
    \includegraphics[width=0.9\textwidth]{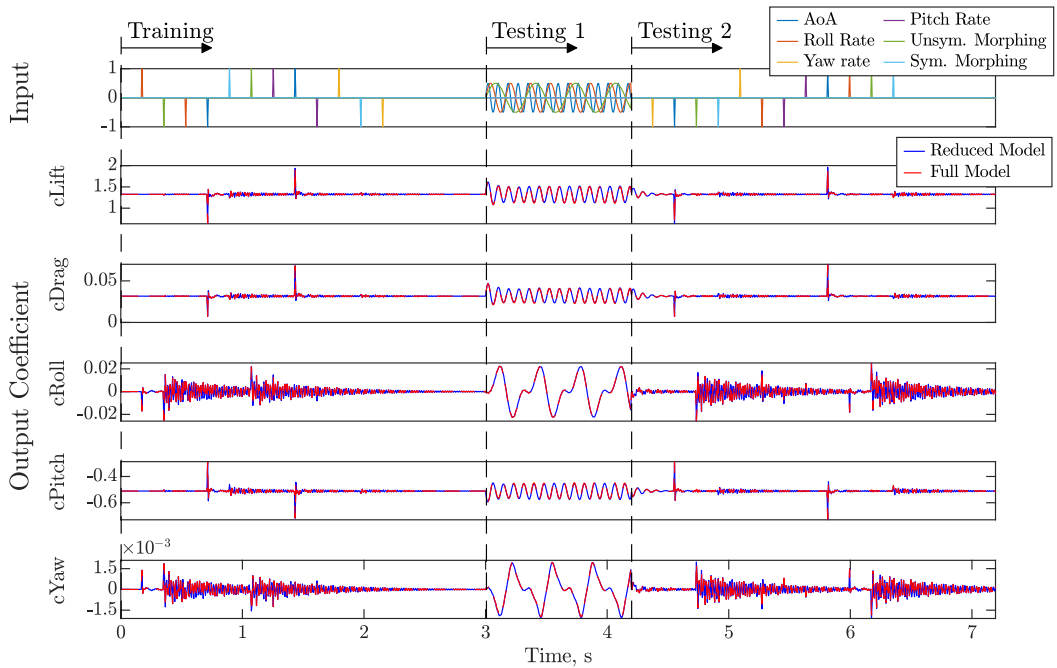}
        \vspace{-.1in}
  \caption{Morphing AWE wing: training phase followed by testing phases, comparing the full and aDMDc models.}
  \label{fig_3}
  \end{center}
      \vspace{-.1in}
\end{figure}

The aeroelastic responses are shown for all three phases in figure~\ref{fig_3}.  In all cases, the agreement between the full and  reduced-order model is excellent. 
Table~\ref{table_example2} summarizes the speed-up factor between the full model and the ROM and the relative errors in the predicted response.  

\begin{table}[b]
\vspace{-.05in}
\caption{Morphing wing ROM speed-up factor $S$ and mean relative errors $R_i$.}\label{table_example2}
\begin{center}
\vspace{-.25in}
\begin{tabular}{lllllll}
\hline
Input: & $S$ & $R_{Lift}$ & $R_{Drag}$ & $R_{Roll}$ & $R_{Pitch}$ & $R_{Yaw}$ \\
\hline
1. Reference: & 2894 & 0.84\% & 1.81\% & 1.20\% & 0.88\% & 3.13\% \\
2. Random: & 2894 & 0.52\% & 1.12\% & 4.95\% & 0.59\% & 5.59\% \\
\hline
\end{tabular}
\vspace*{-4pt}
\end{center}
\vspace{-.3in}
\end{table}

\subsection{Parametric reduced-order aeroelastic model}

Large changes in operating conditions generally pose a challenge for ROMs. Indeed, when the operating condition changes enough, the ROM loses its validity and another training phase must be performed.  In this example, we demonstrate the ability to interpolate several ROMs that are identified at different speeds.  The result is a highly flexible model that seamlessly covers a large range of flight conditions and results in high performance for continuously varying operating conditions. We only need to interpolate models that are identified for different flow velocities, as the other parameter variations, such as angle of attack, are captured well by a single ROM for a given velocity. 

\begin{figure}[t]
\vspace{-.3in}
\begin{center}
    \includegraphics[width=0.9\textwidth]{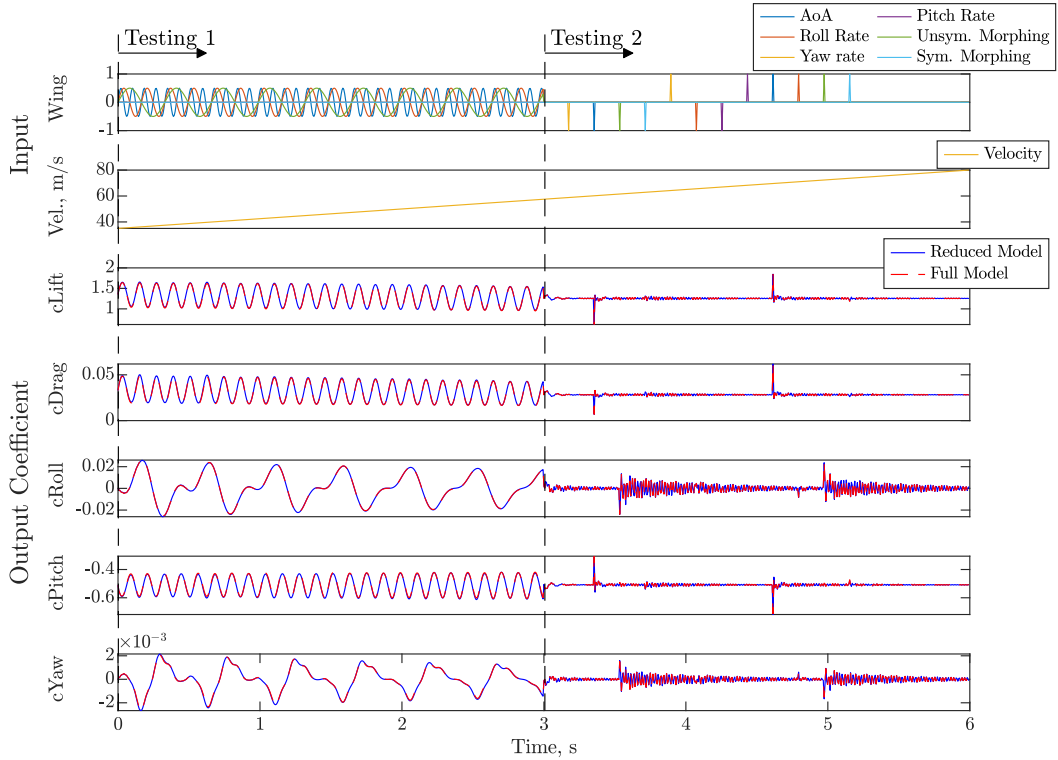}
  \caption{Morphing AWE wing: reference and random testing inputs with linear velocity sweep from $V_{\infty} = 35\ m/s$ to $V_{\infty} = 80\ m/s$, comparing the full and parametric aDMDc models.}
  \label{fig_4}
  \end{center}
\end{figure}

A set of five ROMs are generated for the ftero wing, following the scheme presented above, for velocities in the range $V_\infty = 35 - 80$m/s. These ROMs are then used to predict the behaviour of the system using the reference set of inputs, with a linear velocity sweep across the entire range. The same inputs are also used for the full model, and the results are presented and compared in figure~\ref{fig_4}.

\begin{table}[b]
\caption{Speed-up factor $S$ and mean relative errors $R_i$ for parametric ROM.}\label{table_example3}
\begin{center}
\vspace{-.1in}
\begin{tabular}{lllllll}
\hline
Input: & $S$ & $R_{Lift}$ & $R_{Drag}$ & $R_{Roll}$ & $R_{Pitch}$ & $R_{Yaw}$ \\
\hline
1. Reference: & 1338 & 0.89\% & 1.82\% & 1.67\% & 0.98\% & 3.92\% \\
2. Random: & 1338 & 0.29\% & 0.59\% & 3.90\% & 0.27\% & 
4.83\% \\
\hline
\end{tabular}
\vspace*{-4pt}
\end{center}
\vspace{-.3in}
\end{table}

Again, there is excellent agreement between the full-order and reduced-order models.  
 Furthermore, even after the interpolation is introduced where multiple systems are solved in parallel, the ROMs are still more than three orders of magnitude faster than the full model.  These results are summarized in Tab.~\ref{table_example3}.

\subsection{Model predictive control}

To further demonstrate the performance of the AWE morphing wing ROM, the model is incorporated into an MPC framework. The MPC is run on three test cases incorporating different levels of fidelity of the aDMDc ROM. First, the MPC architecture is demonstrated on a constant flight speed gust load alleviation case, based on an aDMDc model in a single flight condition. 
Next, a constant flight speed lift-force tracking MPC demonstrates the use of the aDMDc model with nonlinear calculation of the aerodynamic forces. 
Finally, the parametric aDMDc model is incorporated into the MPC framework, tracking the objective of a desired lift coefficient while flying large-amplitude AWE trajectories over a range of operating conditions and in gusty wind fields. To assess an accurate response of the morphing wing, the MPC is tested using the full high fidelity aeroelastic model. For simplicity, Matlab's interior-point method fmincon is used as the optimization routine for MPC. Additional computational acceleration may be achieved by using MPC specific optimization routines in the future. In all three test cases, the current state vector is available for feedback, bypassing sensor selection and observer filter design.

\paragraph{Gust load alleviation MPC.}

Maintaining the wing stress level below a critical threshold is crucial to guarantee structural integrity of the wing. The stress level in the wing is directly proportional to the modal amplitudes of the wing structure, especially to the first bending mode~\cite{Haghighat2012}. Therefore, the MPC objective in this first example is to keep the first bending mode amplitude close to zero, ensuring structural integrity of the wing at gust encounter. This goal is achieved by actuating the morphing wings symmetrically, thereby adjusting the lift and the wing root bending moment. The first bending mode is available for feedback in this study, which could be estimated for example with strain gauges placed at the wing root. Flying at a constant velocity of $V = 60m/s$ and with the first bending mode being a physical state of the aDMDc model, the nonlinear calculation of the aerodynamic forces and moments can be omitted, resulting in linear model. The system time step is $\Delta t_{sys} = 0.006s$ and the MPC time step is $\Delta t_{model} = 0.018s$, representing approximately a state-of-the-art flight control system actuator update frequency~\cite{Meier2015}. The weighting matrices are $Q = 10$, $R_u = 0$, and $R_{\Delta u} = 1$, the actuation is limited to $u \in [-1,1]$, $\Delta u \in [-0.18,0.18]$, and the control and prediction horizon are set to $m_p = m_c = 10$. 

\begin{figure}[t]
\centering\includegraphics[width=5.5in]{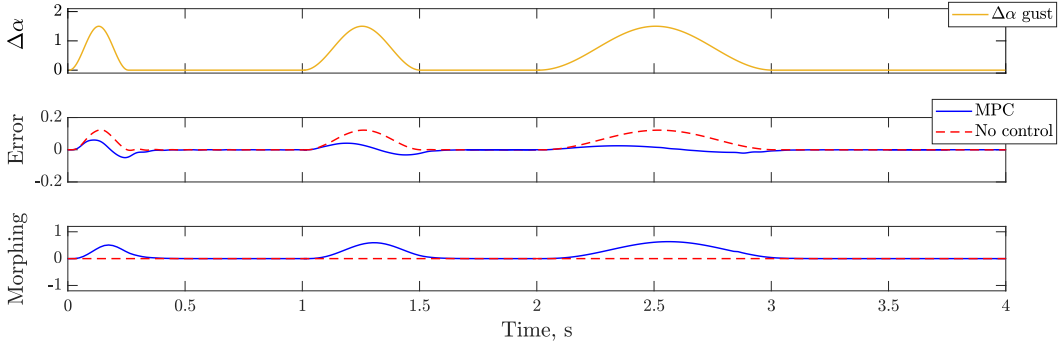}
\vspace{-.05in}
\caption{MPC gust load alleviation using morphing wing. Gust induced angle of attack: \hspace{\textwidth} $\Delta\alpha=1.5^o$, $t_g = [0.25s, 0.5s, 1s]$. Error: first bending mode amplitude deviation.}
\label{fig_MPC1}
\vspace{-.1in}
\end{figure}

The wing encounters a series of three gusts with constant amplitude $\Delta \alpha = 1.5^o$ and increasing gust length $t_{g,1} = 0.25s, t_{g,2} = 0.5s,$ and $t_{g,3} = 1.0s$. Figure \ref{fig_MPC1} shows the gust induced angle of attack in the upper plot, the modal amplitude deviation of the first bending mode calculated with the high fidelity model in the middle plot, and the actuation of the morphing wing in the lower plot. First, the MPC is switched on (blue continuous line) and second, the MPC is switched off (red dashed line). The expected delay of the MPC is clearly visible, especially prominent for short gust lengths, which could be dealt with using additional gust-induced angle of attack sensors~\cite{Haghighat2012}. Nevertheless, the performance of the controller is promising, achieving modal amplitude peak reductions of $\Delta q_{0.25s} = 51.2\%$ $\Delta q_{0.5s} = 66.5\%$, and $\Delta q_{1.0s} = 79.5\%$.

\paragraph{Lift force tracking MPC.}

In the second test case, the MPC objective is to track a specified lift coefficient. We thus use the full aDMDc ROM with the nonlinear calculation of the lift. The lift is directly available for feedback in this study, which could be measured on the drone for example by a five-hole probe~\cite{Hesse2016}. Tracking a specific lift is similarly achieved by symmetrically actuating the morphing wing to reduce or increase the lift coefficient. 
\begin{figure}[t]
\centering\includegraphics[width=5.5in]{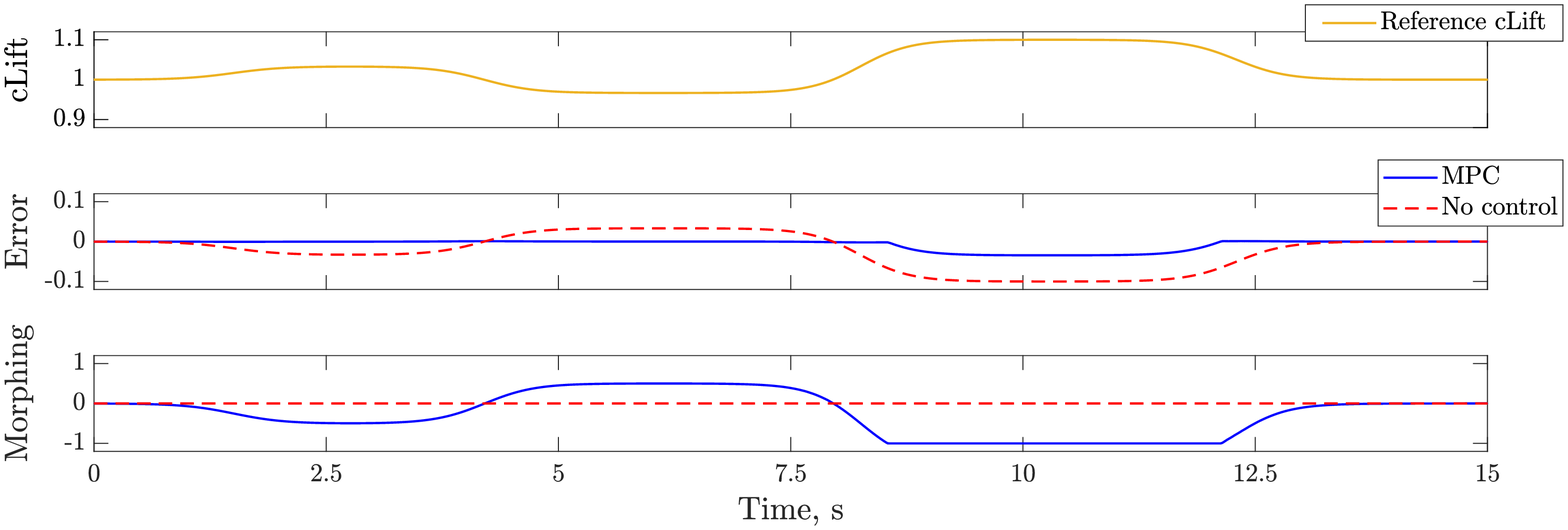}
\vspace{-.05in}
\caption{MPC lift tracking using morphing wings.}
\label{fig_MPC2}
\vspace{-.2in}
\end{figure}
The MPC parameters are the same as in the first example, except the weighting matrices are set to $Q = 1$ and $R_{\Delta u} = 10000$. Figure \ref{fig_MPC2} shows the desired normalized lift coefficient in the top plot, the error defined as the deviation of the actual from the desired normalized lift assessed by the high fidelity model in the middle plot, and the morphing actuation in the bottom plot. Again, the MPC is first switched on (blue continuous line), and then switched off (red dashed line). The MPC performs well, even when the commanded lift exceeds the attainable lift and the actuation is saturated.

\begin{figure}[t]
\centering\includegraphics[width=5.5in]{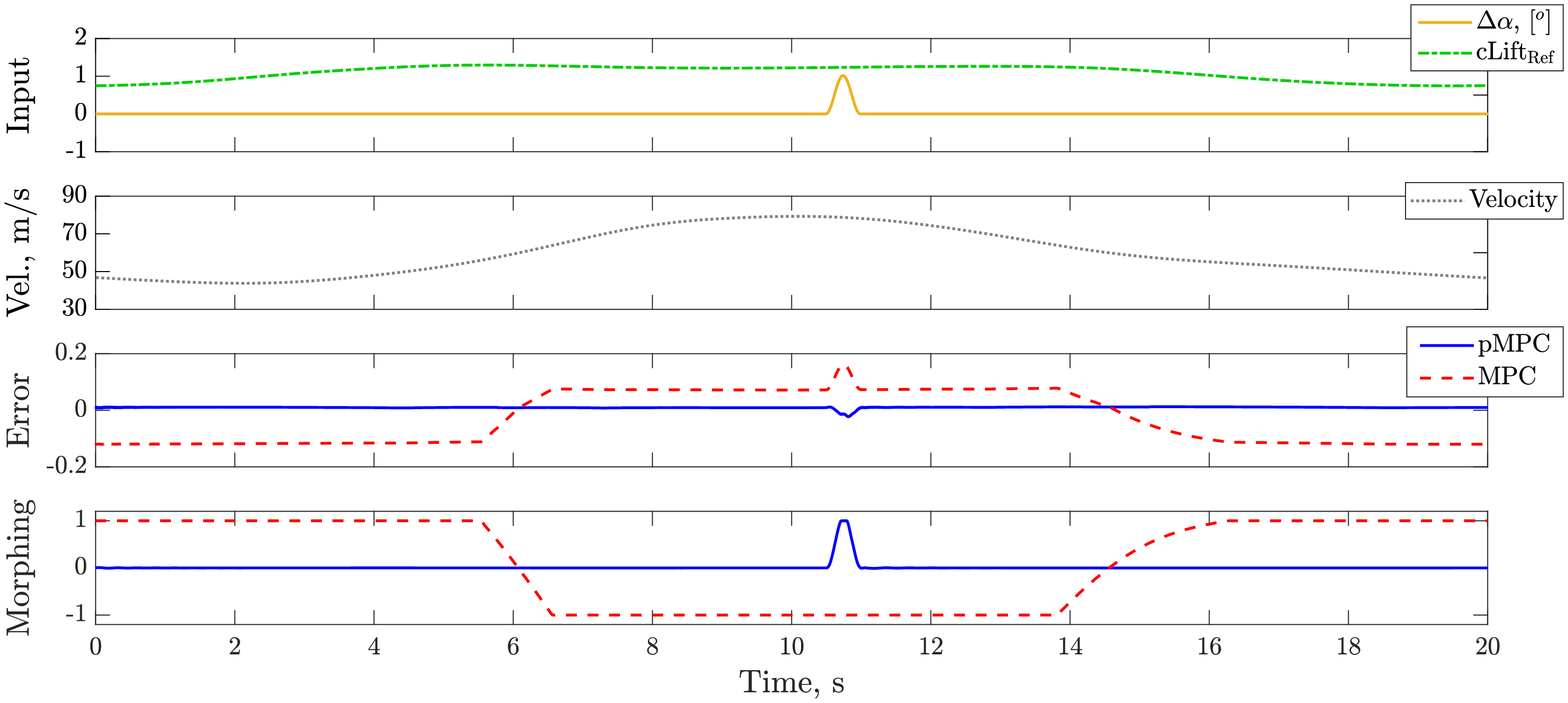}
\vspace{-.05in}
\caption{Parametric MPC (pMPC) vs. non-parametric MPC (MPC) load alleviation using morphing for large flight speed range. The controller tracks a reference cLift by symmetrically morphing the left and right wing in a typical AWE power cycle, where a vertical gust hits the wing at maximum flight speed.}
\label{fig_MPC3}
\end{figure}

\paragraph{AWE trajectory and gust load alleviation MPC.} 
The third test case is AWE specific, where the structural integrity of the tether is maintained by keeping the tether force below a critical maximum force. 
Currently, the tether force is controlled by changing the ground station reel out speed; however, this approach is too slow to mitigate force peaks introduced by gusts on the wing. 
Therefore, either the maximum allowable tether force must be reduced, leading to lower power production, or active load alleviation must be achieved on the wing. 
In this example, we use MPC to keep the lift coefficient on the wing close to a desired steady state value. 
The drone flies an AWE specific trajectory, with flight velocities ranging between $V = [40, 80]m/s$, shown in figure~\ref{fig_MPC3}. Additionally, the wing encounters a gust at the maximum flight speed, with $t_g = 0.5s$ and $\Delta \alpha = 1^o$. 
In this case, an AWE specific lift coefficient trajectory is defined, with a higher lift coefficient in the traction phase at high flight speeds and lower lift coefficient at low flight speeds in the retraction phase. This is achieved by both changing the drone's angle of attack for large and slow lift coefficient changes (assumed to be controlled by an extended fixed wing drone controller using the drone elevator~\cite{Fasel2019rom}) and by symmetrically actuating the morphing wing for fast gust load alleviation. To accurately control the system over the large range of flight speeds, the parametric aDMDc-based ROM is incorporated in the MPC. As the objective of the MPC is similar to case 2, the MPC parameters are the same as in the previous example. 

The results of this test case are shown in figure~\ref{fig_MPC3}. We compare the MPC performance with the parametric aDMDc model (blue continuous line), and the MPC performance with a non-parametric aDMDc ROM obtained at the mean flight speed and assuming a quadratic depencency of the lift force on the flight velocity (red dashed line). 
Thus, nonlinearities introduced by the flexibility of the morphing wing and by the reduced frequency of the unsteady aerodynamics are neglected. The results highlight the importance of using a parametric ROM covering the full operational regime to precisely track a desired lift coefficient when dealing with large flight speed ranges specific to AWE. The non-parametric ROM predicts the lift coefficient inaccurately and therefore saturates the actuator. Therefore, it not only fails to track the correct lift coefficient but also prevents any gust load alleviation. This could potentially lead to tether rupture and loss of drone, further motivating the parametric ROM.

\section{Discussion and conclusion}

In this work, we present a data-driven reduced-order modeling framework for flexible aerospace structures that is valid over a large range of operating conditions. Our models are based on the recent DMDc algorithm, and we introduce two methodological extensions to this algorithm to broaden its applicability. First, we generalize the formulation to handle algebraic equations, and second, we develop an interpolation scheme to smoothly connect several models valid in different operating regimes. The method is demonstrated on several test cases, ranging from a generic differential-algebraic system of equations to a high fidelity, three-dimensional numerical model of a morphing wing AWE system. The reduced-order models are faster than real time and are able to accurately predict nonlinear, unsteady aeroelastic responses. Furthermore, the model is incorporated into an MPC framework, demonstrating the potential to perform load control of morphing wings in AWE typical flight conditions, characterized by large changes to the load and flight speed. 

The results of this study show that our extension to the DMDc algorithm to handle algebraic equations is vital for unsteady aerodynamic and aeroelastic systems, not only to capture the correct dynamics but especially to retain stability of the system. The ROM shows excellent agreement with the full-order model for unsteady aerodynamic lift, drag, and moment coefficient prediction, with relative errors below $0.5\%$. For the coupled aeroelastic system, the agreement is similarly close, even for large random inputs, with maximum errors for lift and drag below $2\%$ and for the moment coefficients below $6\%$. 
Further, we show that it is possible to smoothly interpolate between several aDMDc ROMs, resulting in a parametric ROM that is valid over a large range of operating conditions and that agrees closely with the full model.  The errors are below $2\%$ for the lift and drag and below $5\%$ for the moment coefficients. In all cases, the ROMs are more than three orders of magnitude faster than the full model, enabling use for MPC. On three test cases, the model performance is evaluated, showing excellent results when performing load control of the morphing wing AWE system. Especially for the AWE specific flight trajectory with gust rejection, the parametric ROM is able to fully mitigate the gust induced load peak, which is not achievable by a single non-parametric ROM.

Although the data-driven modeling procedure introduced in this work is generally applicable, it was specifically motivated by AWE applications.  The resulting ROMs will improve the prediction performance of current AWE models, which are mostly based on simplified quasi-steady aerodynamic models. This will lead to better performing AWE drones with enhanced power production capabilities and potentially reduce the levelized cost of energy, crucial in adopting the emerging AWE technology. Furthermore, incorporating the ROM into existing AWE-specific MPC frameworks with power production objectives~\cite{Stastny2019}, and enhancing them with additional structural stress level constraints, could potentially lead to drastic improvements in control performance and therefore power production. Apart from AWE specific applications of the model, this procedure may be applied to any flexible aerospace structure operating over multiple operating regimes, and more generally to any parametric dynamical system across multiple operating conditions.

\vspace{0.2in}

%%%%%%%%%%%
%% BIBLIOGRAPHY
%%%%%%%%%%%
%\bibliographystyle{ieeetr}
 \begin{spacing}{.9}
 \small{
 \setlength{\bibsep}{6.5pt}
 \bibliographystyle{RS}
 \bibliography{references}

\begin{thebibliography}{99}

\bibitem{Li2018}
Li D, Zhao S, Ronch AD, Xiang J, Drofelnik J, Li Y, Zhang L, Wu Y, Kintscher M,
  Peter H, Rudenko A, Guo S, Yin W, Kirn J, Storm S, Breuker RD. 2018  {A
  review of modelling and analysis of morphing wings Daochun}. {\em Progress in
  Aerospace Sciences} \textbf{100}, 46--62.

\bibitem{Combes:2001}
Combes SA, Daniel TL. 2001  Shape, flapping and flexion: wing and fin design
  for forward flight. {\em The Journal of Experimental Biology} \textbf{204},
  2073--2085.

\bibitem{Birch2001nature}
Birch J, Dickinson M. 2001  Spanwise flow and the attachment of the
  leading-edge vortex on insect wings. {\em Nature} \textbf{412}, 729--733.

\bibitem{Hedenstrom2007science}
Hedenstr{\"o}m A, Johansson L, Wolf M, Von~Busse R, Winter Y, Spedding G. 2007
  Bat flight generates complex aerodynamic tracks. {\em Science} \textbf{316},
  894--897.

\bibitem{song2008aeromechanics}
Song A, Tian X, Israeli E, Galvao R, Bishop K, Swartz S, Breuer K. 2008
  Aeromechanics of membrane wings with implications for animal flight. {\em
  AIAA journal} \textbf{46}, 2096--2106.

\bibitem{leishman:06}
Leishman JG. 2006 {\em Principles of Helicopter Aerodynamics}.
Cambridge, England: Cambridge University Press 2 edition.

\bibitem{dowell:2001}
Dowell EH, Hall KC. 2001  Modeling of Fluid-Structure interaction. {\em Annual
  Review of Fluid Mechanics} \textbf{33}, 445--490.

\bibitem{Proctor2016siads}
Proctor JL, Brunton SL, Kutz JN. 2016  Dynamic mode decomposition with control.
  {\em SIAM Journal on Applied Dynamical Systems} \textbf{15}, 142--161.

\bibitem{Ahrens2014}
Ahrens U, Diehl M, Schmehl R. 2014 {\em {Airborne Wind Energy}}.
Springer.

\bibitem{Cherubini2015}
Cherubini A, Papini A, Vertechy R, Fontana M. 2015  {Airborne Wind Energy
  Systems : A review of the technologies}. {\em Renewable and Sustainable
  Energy Reviews} \textbf{51}, 1461--1476.

\bibitem{fasel2017}
Fasel U, Keidel D, Molinari G, Ermanni P. 2017  Aerostructural optimization of
  a morphing wing for airborne wind energy applications. {\em Smart Materials
  and Structures} \textbf{26}.

\bibitem{Hodges2002}
Hodges DH, Pierce GA. 2002 {\em {Introduction to Structural Dynamics and
  Aeroelasticity}}.
Cambridge University Press.

\bibitem{Molinari2014}
Molinari G, Arrieta AF, Ermanni P. 2014  {Aero-Structural Optimization of
  Three-Dimensional Adaptive Wings with Embedded Smart Actuators}. {\em AIAA
  Journal} \textbf{52}, 1940--1951.

\bibitem{wagner:25}
Wagner H. 1925  {\"Uber} die {Entstehung} des dynamischen {Auftriebes} von
  {Tragfl\"ugeln}. {\em Zeitschrift f\"ur Angewandte Mathematic und Mechanik}
  \textbf{5}, 17--35.

\bibitem{Theodorsen:35}
Theodorsen T. 1935  General theory of aerodynamic instability and the mechanism
  of flutter. Technical Report 496 NACA.

\bibitem{karman:38num}
von Karman T, Sears WR. 1938  Airfoil theory for non-uniform motion. {\em
  Journal Aeronautical Sciences} \textbf{5}, 379--390.

\bibitem{sears:41foil}
Sears WR. 1941  Some aspects of non-stationary airfoil theory and its practical
  applications. {\em AIAA Journal Special Supplement: Centennial of Powered
  Flight} \textbf{8}, 104--108.

\bibitem{peters:2007}
Peters DA, Hsieh MA, Torrero A. 2007  A state-space airloads theory for
  flexible airfoils. {\em Journal of the American Helicopter Society}
  \textbf{52}, 329--342.

\bibitem{Patil1998}
Patil MJ, Hodges DH, Cesnik CE. 1998  {Nonlinear Aeroelastic Analysis of
  Aircraft with High-Aspect-Ratio Wings}. In {\em Proceedings of the 39th
  Structures, Structural Dynamics, and Materials Conference, AIAA} pp.
  2056--2068.

\bibitem{palacios2010}
Palacios R, Murua J, Cook R. 2010  Structural and aerodynamic models in
  nonlinear flight dynamics of very flexible aircraft. {\em AIAA journal}
  \textbf{48}, 2648--2659.

\bibitem{Afonso2017}
Afonso F, Vale J, Oliveira {\'{E}}, Lau F, Suleman A. 2017  {A review on
  non-linear aeroelasticity of high aspect-ratio wings}. {\em Progress in
  Aerospace Sciences} \textbf{89}, 40--57.

\bibitem{Kutz2016book}
Kutz JN, Brunton SL, Brunton BW, Proctor JL. 2016 {\em Dynamic Mode
  Decomposition: Data-Driven Modeling of Complex Systems}.
SIAM.

\bibitem{brunton2019}
Brunton SL, Kutz JN. 2019 {\em Data-driven Science and Engineering: Machine
  Learning, Dynamical Systems, and Control}.
Cambridge University Press.

\bibitem{Brunton2020arfm}
Brunton SL, Noack BR, Koumoutsakos P. 2020  Machine Learning for Fluid
  Mechanics. {\em Annual Review of Fluid Mechanics} \textbf{52}.

\bibitem{Schmid2010jfm}
Schmid PJ. 2010  Dynamic Mode Decomposition of Numerical and Experimental Data.
  {\em Journal of Fluid Mechanics} \textbf{656}, 5--28.

\bibitem{Rowley2009jfm}
Rowley CW, Mezi\'c I, Bagheri S, Schlatter P, Henningson D. 2009  Spectral
  analysis of nonlinear flows. {\em Journal of Fluid Mechanics} \textbf{645},
  115--127.

\bibitem{Tu2014jcd}
Tu JH, Rowley CW, Luchtenburg DM, Brunton SL, Kutz JN. 2014  On dynamic mode
  decomposition: Theory and applications. {\em Journal of Computational
  Dynamics} \textbf{1}, 391--421.

\bibitem{Besselink2013}
Besselink B, Tabak U, Lutowska A, {Van De Wouw} N, Nijmeijer H, Rixen DJ,
  Hochstenbach ME, Schilders WH. 2013  {A comparison of model reduction
  techniques from structural dynamics, numerical mathematics and systems and
  control}. {\em Journal of Sound and Vibration} \textbf{332}, 4403--4422.

\bibitem{Fasel2019rom}
Fasel U, Tiso P, Keidel D, Molinari G, Ermanni P. 2019  {Reduced-Order Dynamic
  Model of a Morphing Airborne Wind Energy Aircraft}. {\em AIAA Journal}
  \textbf{57}, 3586--3598.

\bibitem{katz2001}
Katz J, Plotkin A. 2001 {\em Low-speed aerodynamics} vol.~13.
Cambridge university press.

\bibitem{ftero}
Affentranger L, Baumann L, Canonica R, Gehri I, K{\"o}nig G, Mattia C,
  Michalski A, Wiesem{\"u}ller F, Wild O, Fasel U, Keidel D, Molinari G,
  Ermanni P. 2017  ftero-On the Development of an Airborne Wind Energy System.
  In {\em Airborne Wind Energy Conference 2017}.

\bibitem{keidel2017ftero}
Keidel D, Fasel U, Baumann L, Molinari G, Ermanni P. 2017a  Experimental
  Validation of a Morphing Wing for Airborne Wind Energy Applications. In {\em
  28th International Conference on Adaptive Structures and Technologies,
  Cracow, Poland}.

\bibitem{Keidel2017smasis}
Keidel D, Fasel U, Molinari G, Ermanni P. 2017b  {Design, Development, and
  Structural testing of a Camber-Morphing Flying Wing Airplane}. In {\em
  Conference on Smart Materials, Adaptive Structures and Intelligent Systems}
  Snowbird, UT, USA.

\bibitem{Keidel2020}
Keidel D, Fasel U, Ermanni P. 2020  {Control Authority of a Camber Morphing
  Flying Wing}. {\em Journal of Aircraft, in press}.

\bibitem{fung2008}
Fung YC. 2008 {\em An introduction to the theory of aeroelasticity}.
Courier Dover Publications.

\bibitem{rodden1994}
Rodden WP, Johnson EH. 1994 {\em MSC/NASTRAN aeroelastic analysis: user's
  guide; Version 68}.
MacNeal-Schwendler Corporation.

\bibitem{kier2005}
Kier T. 2005  Comparison of unsteady aerodynamic modelling methodologies with
  respect to flight loads analysis. In {\em AIAA Atmospheric Flight Mechanics
  Conference and Exhibit} p. 6027.

\bibitem{albano1969}
Albano E, Rodden WP. 1969  A doublet-lattice method for calculating lift
  distributions on oscillating surfaces in subsonic flows. {\em AIAA journal}
  \textbf{7}, 279--285.

\bibitem{falkner1943}
Falkner VM. 1943  The calculation of aerodynamic loading on surfaces of any
  shape. Technical report Aeronautical Research Council London (United
  Kingdom).

\bibitem{hedman1965}
Hedman S. 1965  Vortex lattice method for calculation of quasi steady state
  loadings on thin elastic wings. {\em Aeronautical Research Institute of
  Sweden Rept} \textbf{105}.

\bibitem{maute2001}
Maute K, Nikbay M, Farhat C. 2001  Coupled analytical sensitivity analysis and
  optimization of three-dimensional nonlinear aeroelastic systems. {\em AIAA
  journal} \textbf{39}, 2051--2061.

\bibitem{morino1974}
Morino L. 1974  A general theory of unsteady compressible potential
  aerodynamics. Technical Report CR-2464 NASA.

\bibitem{Brunton2013jfs}
Brunton SL, Rowley CW. 2013  Empirical state-space representations for
  {T}heodorsen's lift model. {\em Journal of Fluids and Structures}
  \textbf{38}, 174--186.

\bibitem{Brunton2013jfm}
Brunton SL, Rowley CW, Williams DR. 2013  Reduced-order unsteady aerodynamic
  models at low {R}eynolds numbers. {\em Journal of Fluid Mechanics}
  \textbf{724}, 203--233.

\bibitem{Brunton2014jfs}
Brunton SL, Dawson ST, Rowley CW. 2014  State-space model identification and
  feedback control of unsteady aerodynamic forces. {\em Journal of Fluids and
  Structures} \textbf{50}, 253--270.

\bibitem{drela1999}
Drela M. 1999  Integrated simulation model for preliminary aerodynamic,
  structural, and control-law design of aircraft. In {\em 40th Structures,
  Structural Dynamics, and Materials Conference and Exhibit} p. 1394.

\bibitem{murua2012}
Murua J, Palacios R, Graham JMR. 2012  Applications of the unsteady
  vortex-lattice method in aircraft aeroelasticity and flight dynamics. {\em
  Progress in Aerospace Sciences} \textbf{55}, 46--72.

\bibitem{lieu2005}
Lieu T, Farhat C, Lesoinne M. 2005  POD-based aeroelastic analysis of a
  complete F-16 configuration: ROM adaptation and demonstration. In {\em 46th
  AIAA/ASME/ASCE/AHS/ASC Structures, Structural Dynamics and Materials
  Conference} p. 2295.

\bibitem{amsallem2008}
Amsallem D, Farhat C. 2008  Interpolation method for adapting reduced-order
  models and application to aeroelasticity. {\em AIAA journal} \textbf{46},
  1803--1813.

\bibitem{silva2001}
Silva W, Raveh D. 2001  Development of unsteady aerodynamic state-space models
  from CFD-based pulse responses. In {\em 19th AIAA Applied Aerodynamics
  Conference} p. 1213.

\bibitem{guendel2001}
Guendel R, Cesnik C. 2001  Aerodynamic impulse response of a panel method. In
  {\em 19th AIAA Applied Aerodynamics Conference} p. 1210.

\bibitem{silva2001.2}
Silva WA, Beran PS, Cesnik CE, Guendel RE, Kurdila A, Prazenica RJ, Librescu L,
  Marzocca P, Raveh DE. 2001  Reduced-order modeling: Cooperative research and
  development at the NASA Langley Research Center. In {\em International Forum
  on Aeroelasticity and Structural Dynamics IFASD 2001-008}.

\bibitem{raveh2001}
Raveh DE. 2001  Reduced-order models for nonlinear unsteady aerodynamics. {\em
  AIAA journal} \textbf{39}, 1417--1429.

\bibitem{Brockett1976automatica}
Brockett RW. 1976  Volterra series and geometric control theory. {\em
  Automatica} \textbf{12}, 167--176.

\bibitem{juang1985}
Juang JN, Pappa RS. 1985  An eigensystem realization algorithm for modal
  parameter identification and model reduction. {\em Journal of guidance,
  control, and dynamics} \textbf{8}, 620--627.

\bibitem{Brunton2016pnas}
Brunton SL, Proctor JL, Kutz JN. 2016  Discovering governing equations from
  data by sparse identification of nonlinear dynamical systems. {\em
  Proceedings of the National Academy of Sciences} \textbf{113}, 3932--3937.

\bibitem{Loiseau2017jfm}
Loiseau JC, Brunton SL. 2018  Constrained Sparse {Galerkin} Regression. {\em
  Journal of Fluid Mechanics} \textbf{838}, 42--67.

\bibitem{Loiseau2018jfm}
Loiseau JC, Noack BR, Brunton SL. 2018  Sparse reduced-order modeling:
  Sensor-based dynamics to full-state estimation. {\em Journal of Fluid
  Mechanics} \textbf{844}, 459--490.

\bibitem{Kaiser2018prsa}
Kaiser E, Kutz JN, Brunton SL. 2018  Sparse identification of nonlinear
  dynamics for model predictive control in the low-data limit. {\em Proceedings
  of the Royal Society of London A} \textbf{474}.

\bibitem{de2009}
De~Caigny J, Camino JF, Swevers J. 2009  Interpolating model identification for
  SISO linear parameter-varying systems. {\em Mechanical Systems and Signal
  Processing} \textbf{23}, 2395--2417.

\bibitem{benner2015}
Benner P, Gugercin S, Willcox K. 2015  A survey of projection-based model
  reduction methods for parametric dynamical systems. {\em SIAM review}
  \textbf{57}, 483--531.

\bibitem{baur2017}
Baur U, Benner P, Haasdonk B, Himpe C, Martini I, Ohlberger M. 2017  Comparison
  of methods for parametric model order reduction of time-dependent problems.
  {\em Model Reduction and Approximation: Theory and Algorithms} pp. 377--407.

\bibitem{balas2002}
Balas GJ. 2002  Linear, parameter-varying control and its application to
  aerospace systems. In {\em ICAS congress proceedings}.

\bibitem{Hemati2016aiaa}
Hemati MS, Dawson ST, Rowley CW. 2016  Parameter-Varying Aerodynamics Models
  for Aggressive Pitching-Response Prediction. {\em AIAA Journal} pp. 693--701.

\bibitem{steinbuch2003}
Steinbuch M, van~de Molengraft M, Van Der~Voort AJ. 2003  Experimental
  modelling and LPV control of a motion system. In {\em 2003 American control
  conference, ACC} pp. 1374--1379.

\bibitem{wassink2005}
Wassink MG, van~de Wal M, Scherer C, Bosgra O. 2005  LPV control for a wafer
  stage: beyond the theoretical solution. {\em Control Engineering Practice}
  \textbf{13}, 231--245.

\bibitem{paijmans2008}
Paijmans B, Symens W, Van~Brussel H, Swevers J. 2008  Identification of
  interpolating affine LPV models for mechatronic systems with one varying
  parameter. {\em European Journal of Control} \textbf{14}, 16--29.

\bibitem{camacho2013model}
Camacho EF, Alba CB. 2013 {\em Model predictive control}.
Springer Science \& Business Media.

\bibitem{garcia1989model}
Garcia CE, Prett DM, Morari M. 1989  Model predictive control: theory and
  practice---a survey. {\em Automatica} \textbf{25}, 335--348.

\bibitem{morari1999model}
Morari M, Lee JH. 1999  Model predictive control: past, present and future.
  {\em Computers \& Chemical Engineering} \textbf{23}, 667--682.

\bibitem{mayne2014automatica}
Mayne DQ. 2014  Model predictive control: Recent developments and future
  promise. {\em Automatica} \textbf{50}, 2967--2986.

\bibitem{Doyle2013book}
Doyle JC, Francis BA, Tannenbaum AR. 2013 {\em Feedback control theory}.
Courier Corporation.

\bibitem{Brunton2015amr}
Brunton SL, Noack BR. 2015  Closed-loop turbulence control: Progress and
  challenges. {\em Applied Mechanics Reviews} \textbf{67},
  050801--1--050801--48.

\bibitem{Peng.2009}
Peng H, Wu J, Inoussa G, Deng Q, Nakano K. 2009  Nonlinear system modeling and
  predictive control using the {RBF} nets-based quasi-linear {ARX} model. {\em
  Control Engineering Practice} \textbf{17}, 59--66.

\bibitem{Zhang2016icra}
Zhang T, Kahn G, Levine S, Abbeel P. 2016  Learning deep control policies for
  autonomous aerial vehicles with {MPC}-guided policy search. In {\em IEEE Int
  Conf Robotics Autom} pp. 528--535.

\bibitem{Ilzhofer2007}
Ilzh{\"{o}}fer A, Houska B, Diehl M. 2007  {Nonlinear MPC of kites under
  varying wind conditions for a new class of large-scale wind power
  generators}. {\em International journal of robust and nonlinear control}
  \textbf{17}, 1590--1599.

\bibitem{Gros2013}
Gros S, Zanon M, Diehl M. 2013  {Control of Airborne Wind Energy Systems Based
  on Nonlinear Model Predictive Control {\&} Moving Horizon Estimation}. In
  {\em European Control Conference (ECC)} pp. 1017--1022.

\bibitem{Lucia2014}
Lucia S, Engell S. 2014  {Control of Towing Kites under Uncertainty using
  Robust Economic Nonlinear Model Predictive Control}. In {\em 2014 European
  Control Conference (ECC)} pp. 1158--1163.

\bibitem{Wood2017}
Wood TA, Ahbe E, Hesse H, Smith RS. 2017  {Predictive Guidance Control for
  Autonomous Kites with Input Delay}. {\em IFAC Pap} \textbf{50}, 9--14.

\bibitem{Diwale2017}
Diwale S, Faulwasser T, Jones CN. 2017  {Model Predictive Path-Following
  Control for Airborne Wind Energy Systems}. {\em IFAC Pap} \textbf{50}.

\bibitem{Stastny2019}
Stastny T, Ahbe E, Dangel M, Siegwart R. 2019  {Locally Power-optimal Nonlinear
  Model Predictive Control for Fixed-wing Airborne Wind Energy}. {\em 2019
  American Control Conference (ACC)} pp. 2191--2196.

\bibitem{Haghighat2012}
Haghighat S, Liu HHT, Martins JR. 2012  {Model-Predictive Gust Load Alleviation
  Controller for a Highly Flexible Aircraft}. {\em Journal of Guidance, Control
  and Dynamics} \textbf{35}.

\bibitem{Simpson2014}
Simpson RJS, Palacios R, Hesse H, Goulart P. 2014  {Predictive Control for
  Alleviation of Gust Loads on Very Flexible Aircraft}. In {\em AIAA SciTech
  Forum}.

\bibitem{Wang2016}
Wang Y, Wynn A, Palacios R. 2016  {Model-Predictive Control of Flexible
  Aircraft using Nonlinear Reduced-Order Models}. In {\em AIAA SciTech Forum}.

\bibitem{Wang2018}
Wang Y, Wynn A, Palacios R. 2018  Nonlinear Aeroelastic Control of Very
  Flexible Aircraft Using Model Updating. {\em Journal of Aircraft}
  \textbf{55}, 1551--1563.

\bibitem{Artola2019}
Artola M, Wynn A, Palacios R. 2019  A Nonlinear Modal-Based Framework for Low
  Computational Cost Optimal Control of 3D Very Flexible Structures. In {\em
  18th European Control Conference (ECC)} pp. 3836--3841.

\bibitem{harder1972}
Harder RL, Desmarais RN. 1972  Interpolation using surface splines. {\em
  Journal of Aircraft} \textbf{9}, 189--191.

\bibitem{shepard1968}
Shepard D. 1968  A two-dimensional interpolation function for
  irregularly-spaced data. In {\em Proceedings of the 1968 23rd ACM national
  conference} pp. 517--524. ACM.

\bibitem{newmark1959}
Newmark NM. 1959  A method of computation for structural dynamics. American
  Society of Civil Engineers.

\bibitem{gavish2014}
Gavish M, Donoho DL. 2014  The optimal hard threshold for singular values is
  $4/\sqrt{3} $. {\em IEEE Transactions on Information Theory} \textbf{60},
  5040--5053.

\bibitem{Meier2015}
Meier L, Honegger D, Pollefeys M. 2015  {PX4: A node-based multithreaded open
  source robotics framework for deeply embedded platforms}. In {\em ICRA}.

\bibitem{Hesse2016}
Hesse H, Palacios R. 2016  {Dynamic Load Alleviation in Wake Vortex
  Encounters}. {\em Journal of Guidance, Control, and Dynamics} \textbf{39},
  801--813.

\bibitem{stability2001}
Bhagwat MJ, Leishman JG. 2001  Stability, consistency and convergence of
  time-marching free-vortex rotor wake algorithms. {\em Journal of the American
  Helicopter Society} \textbf{46}, 59--71.

\bibitem{sarpkaya1989}
Sarpkaya T. 1989  Computational methods with vortices---the 1988 Freeman
  scholar lecture. {\em Journal of Fluids Engineering} \textbf{111}, 5--52.

\bibitem{openvogel}
Hazebrouck GA. 2018  OpenVOGEL --- An open source project for aerodynamic
  simulations based in potential flow theory. [Online; accessed 12-April-2019].

\bibitem{widnall1975}
Widnall SE. 1975  The structure and dynamics of vortex filaments. {\em Annual
  Review of Fluid Mechanics} \textbf{7}, 141--165.

\bibitem{bristow1980}
Bristow D. 1980  Development of panel methods for subsonic analysis and design.
  Technical Report 3234 NASA.

\end{thebibliography}
 }
 \end{spacing}

\section{Appendix}

\subsection{Full flexible dynamics modeling} \label{App1}

The structural model for the presented test cases are all based on a combination of linear plate and beam elements. The external skin of the lifting surface and the Voronoi-based internal structure are modeled using plates, while the stringers are modeled with beam elements. The stiffness and mass matrices are obtained with the commercial software Nastran, and later imported in Matlab to be coupled, via thin plate spline (TPS) and inverse distance weighting (IDW) with the aerodynamic model. The TPS is used to obtain the deformation of the aerodynamic mesh due to structural deformation, while the IDW is used to project aerodynamic forces onto the structure. A detailed description of the structural model and the interpolation can be found in~\cite{Fasel2019rom}.

The aerodynamic model is based on a steady 3D panel method, with a combination of doublets and sources, extended to account for unsteady aerodynamics~\cite{katz2001}. The extension of the steady panel method to the unsteady case is straightforward. At each time step, a new row of doublets, after the trailing edge, is shed. This represents the unsteadiness of the flow and the "memory" in the flow itself. All the other wake nodes are then moved, using a Runge-Kutta integration scheme of second order, using the local velocity of the flow. The aerodynamic forces on the surface are computed with the coefficient of pressure on each panel, considering the far field velocity, the induced velocity by the wing itself, and, in the unsteady case, the induced velocity by all the wake panels.

The basic method has been slightly modified for the purpose of this study. Even if the integration of the location of wake nodes is performed with a Runge-Kutta scheme of second order, which is a necessary condition for the stability of the wake~\cite{stability2001}, it was found that the wake was significantly noisy due to numerical errors. In general, it is difficult to distinguish numerical instabilities from physical wake instabilities~\cite{leishman:06}, the difference could be found recalling that the first numerical instability appears with a spatial frequency equal to half the row spacing in the wake itself~\cite{sarpkaya1989}. In our context, a small, stable oscillation, with this frequency was found. In order to reduce it, the induced velocity by the wake is modified in two ways:

\begin{itemize}
    \item First, the doublet in the wake is substituted with a vortex ring, as the expression for the induced velocity, the Biot-Savart equation, is less unstable in this case~\cite{openvogel}.
    \item Second, a Vatistas core model is implemented following~\cite{leishman:06}. This model is required to avoid high, unphysical induced velocities when one wake node gets close to a singularity~\cite{widnall1975}. This, in practise, consists in the premultiplication of the induced velocity by the Biot-Savart rule, which regularise the function close to the vortex line.
\end{itemize}
The modifications are especially important as, due to the wake rollup, wake panels can be significantly distorted increasing the numerical difficulties due to the non-optimal aspect ratio~\cite{bristow1980}.

In the standard unsteady formulation, during the integration in time, the state of the system continuously increases as new rows of the wake sheet are added. This increases the computational time for the integration at each time step and, for this reason, it was decided to truncate the wake. The truncation has been performed far from the body, and a detailed study on the influence of this on the generalised forces has been conducted. It was concluded that, if the truncation point is far enough, the influence on the force can be neglected.

\end{document}